\documentclass{pasj00}
\usepackage{graphicx}

\begin{document}
\SetRunningHead{S. Morita et al.}{Chromospheric Anemone Jets with
  Hinode/SOT and Hida Ca\emissiontype{II} Spectroheliogram}
\Received{2009/07/04} \Accepted{2009/12/14}

\title{Observations of Chromospheric Anemone Jets with Hinode/SOT and
  Hida Ca\emissiontype{II} Spectroheliogram}

\author{Satoshi \textsc{Morita},  Kazunari \textsc{Shibata},  Satoru \textsc{Ueno},
	Kiyoshi \textsc{Ichimoto}, Reizaburo \textsc{Kitai}, and Ken-ichi \textsc{Otsuji}
}
\affil{Kwasan and Hida Observatories, Kyoto University, \\
Kurabashira, Kamitakara, Takayama, Gifu 506-1314}
\email{morita@kwasan.kyoto-u.ac.jp}


%


\maketitle

\begin{abstract}

We present the first simultaneous observations of chromospheric
``anemone'' jets in solar active regions with Hinode/SOT
Ca\emissiontype{II}~H broadband filetergram and Ca\emissiontype{II}~K
spetroheliogram on the Domeless Solar Telescope (DST) at Hida
Observatory.  During the coordinated observation, 9 chromospheric
anemone jets were simultaneously observed with the two instruments.
These observations revealed three important features, i.e.: (1) the
jets are generated in the lower chromosphere, i.e. these cannot be
seen in Ca\emissiontype{II} K$_{3}$, (2) the length and lifetime of
the jets are 0.4--5 Mm and 40--320 sec, (3) the apparent velocity of
the jets with Hinode/SOT are 3--24 km/s, while Ca\emissiontype{II}
K$_{3}$ component at the jets show blueshifts (in 5 events) in the
range of 2--6 km/s.  The chromospheric anemone jets are associated
with mixed polarity regions which are either small emerging flux
regions or moving magnetic features.  It is found that the
Ca\emissiontype{II} K line often show red or blue asymmetry in
K$_{2}$/K$_{1}$ component: the footpoint of the jets associated with
emerging flux regions often show redshift (2--16 km/s), while the one
with moving magnetic features show blueshift ($\sim5$ km/s).  Detailed
analysis of magnetic evolution of the jet foaming regions revealed
that the reconnection rate (or canceling rate) of the total magnetic
flux at the footpoint of the jets are of order of 10$^{16}$ Mx/s, and
the resulting magnetic energy release rate $(1.1-10)\times10^{24}$
erg/s, with the total energy release $(1-13)\times10^{26}$ erg for the
duration of the magnetic cancellations, $\sim$130s.  These are
comparable to the estimated total energy, $\sim10^{26}$ erg, in a
single chromospheric anemone jet.  In addition to Hida/DST
Ca\emissiontype{II}-K Spectroheliogram and Hinode/SOT
Ca\emissiontype{II}~H broadband filetergram, we also used Hinode/SOT
magnetogram as well as Hida H$\alpha$ filtergram.  An
observation-based physical model of the jet is presented.  The
relation between chromospheric anemone jets and Ellerman bombs is
discussed.
\end{abstract}

\section{Introduction}

The Solar Optical Telescope (SOT) (\cite{tsuneta2008},
\cite{suematsu2008a}) onboard Hinode \citep{kosugi2007} has discovered
ubiquitous tiny jets in the active region chromosphere, called
chromospheric anemone jets, with Ca\emissiontype{II} H broadband
filter observations \citep{shibata2007}. These jets are typically 3--7
arcsec (2--5 Mm) long, and 0.2--0.4 arcsec (0.15--0.3 Mm) wide, and
their apparent velocity is 10--20 km/s. Their morphology shows the
inverted Y shape, which is quite similar to the shape of coronal X-ray
anemone jet discovered by Yohkoh (\cite{shibata1992},
\yearcite{shibata1994}, \cite{shimojo1996}).  Detailed observational
analysis using magnetogram \citep{shimojo1998} and magnetohydrodynamic
numerical simulations (\cite{yokoyama1995}, \yearcite{yokoyama1996})
of the X-ray anemone jet showed that the anemone shape is foamed as a
result of magnetic reconnection, so it can be an indirect
observational evidence of the magnetic reconnection in the solar
corona.  These findings have been recently confirmed and extended to
even smaller X-ray jets with the X-Ray Telescope (XRT) onboard Hinode
(\cite{cirtain2007}, \cite{shimojo2007}, \cite{savcheva2007}).

The discovery of ubiquitous tiny chromospheric anemone jets suggests
that these jets may be generated by the magnetic reconnection similar
to that occurring in the corona.  Recently, \citet{shibata2007}
reported in their preliminary observations that some of the footpoint
of the jets correspond to the mixed magnetic polarities, suggesting
reconnection in the photosphere or chromosphere. The total energy
involved in a single chromospheric anemone jet was estimated to be
10$^{25}$ erg, which is comparable to the energy of nanoflares
proposed ideal by \citet{parker1988}.  Hence it may be interesting to
study the relation between these ubiquitous jets and the coronal
heating, although at present the number of these jets are too small to
explain the coronal heating \citep{shibata2007}.  It should be
stressed that there are even more smaller jets or jet-like features in
the chromosphere whose footpoints are not well resolved so by
definition they cannot be identified as the chromospheric anemone jet.
(Note that a chromospheric anemone jet has a bright footpoint that has
an anemone shape, or inverted Y shape structure.)

There are number of unanswered questions concerning the chromospheric
anemone jet, e.g., What is the true velocity (Doppler velocity) of
these jets?  Are mixed polarities universal at the footpoint of these
jets?  What is the relation to other chromospheric jets, such as
surges (e.g., \cite{rust1968}, \cite{roy1973a}, \cite{kubota1974},
\cite{schmieder1995}, \cite{canfield1996}, \cite{liu2004},
\cite{brooks2007}), H$\alpha$ jets (e.g., \cite{chae1999}), EUV jets
(e.g., \cite{brueckner1983}, \cite{alexander1999}), and spicules
(e.g., \cite{beckers1972}, \cite{nishikawa1988}, \cite{suematsu1995},
\cite{sterling2000}, \cite{depontieu2007}, \cite{suematsu2008b})?  The
footpoint of chromospheric anemone jets reminds us of similar tiny
brightening features, called Ellerman bombs (e.g.,
\cite{ellerman1917}, \cite{roy1973b}, \cite{kurokawa1982},
\cite{kitai1983}, \cite{nindos1998}, \cite{qiu2000},
\cite{geolgoulis2002}, \cite{pariat2004}, \cite{fang2006},
\cite{pariat2007}, \cite{matsumoto2008a}, \yearcite{matsumoto2008b},
\cite{watanabe2008}).  So the question arises whether the footpoints
of these jets correspond to the Ellerman bombs.

In this paper, we report the first simultaneous observations of the
chromospheric anemone jets with SOT/Hinode Ca\emissiontype{II} H
broadband filter and Ca\emissiontype{II} K spectroheliogram on the
Domeless Solar Telescope (DST) at Hida Observatory \citep{nakai1985}.
Since the Hinode/Ca\emissiontype{II} H filter is a broadband filter,
it is not possible to derive velocity and other information such as
occurrence height of the jets. Using the spectroheliogram on DST, we
first derived the true Doppler velocity and height information of the
chromospheric anemone jet. In this paper, we have made comprehensive
analysis of the chromospheric anemone jets, also using SOT/Hinode
magnetogram data as well as Hida H$\alpha$ filtergram, to develop an
observation-based physical model of the jet.

The paper is organized in the following manner: In section 2, we
present observational method for both Hinode and Hida observations,
and in section 3, we describe observational results for typical three
examples of the chromospheric anemone jets in very detail. Finally, we
discuss the energy release rate and the physical model of the jets.

\section{Observation}

In order to study evolution and dynamics of the chromospheric anemone
jets and the related phenomena, we performed coordinated observations
of the active region, using the SOT/Hinode Broadband Filter Imager
(BFI) of Filtergraph (FG) and the 60~cm Domeless Solar Telescope of
the Hida Observatory, Kyoto University. These observations were
performed under the Hinode Observation Plan 12 (HOP12).  The most
important purpose of HOP12 was to get the simultaneous observations of
dynamic phenomena such as jets with the Hinode Ca\emissiontype{II}~H
BFI as well as DST/Hida Ca\emissiontype{II}~K spectroheliograph.

The target region in this study was NOAA AR 10966 and its surrounding
area.  The time of the central meridian passing through the active
region was around 11:00 UT on August~9, 2007.  For the present study,
we used the data set covering the period from 23:00 UT on August~6 to
01:00 UT on August~10, 2007.  The total observing time of the data set
was approximately 25 hours.  The angle between the line-of-sight and
the vector normal to the horizontal plane at the target region ranged
from 36 to 13 degrees (between 0.81 and 0.97 in $\mu=\cos\theta$, see
Table~1), so it was an ``on disk'' observation.

\subsection{Hinode SOT observation}

\begin{figure}
  \begin{center}
    \FigureFile(78mm,78mm){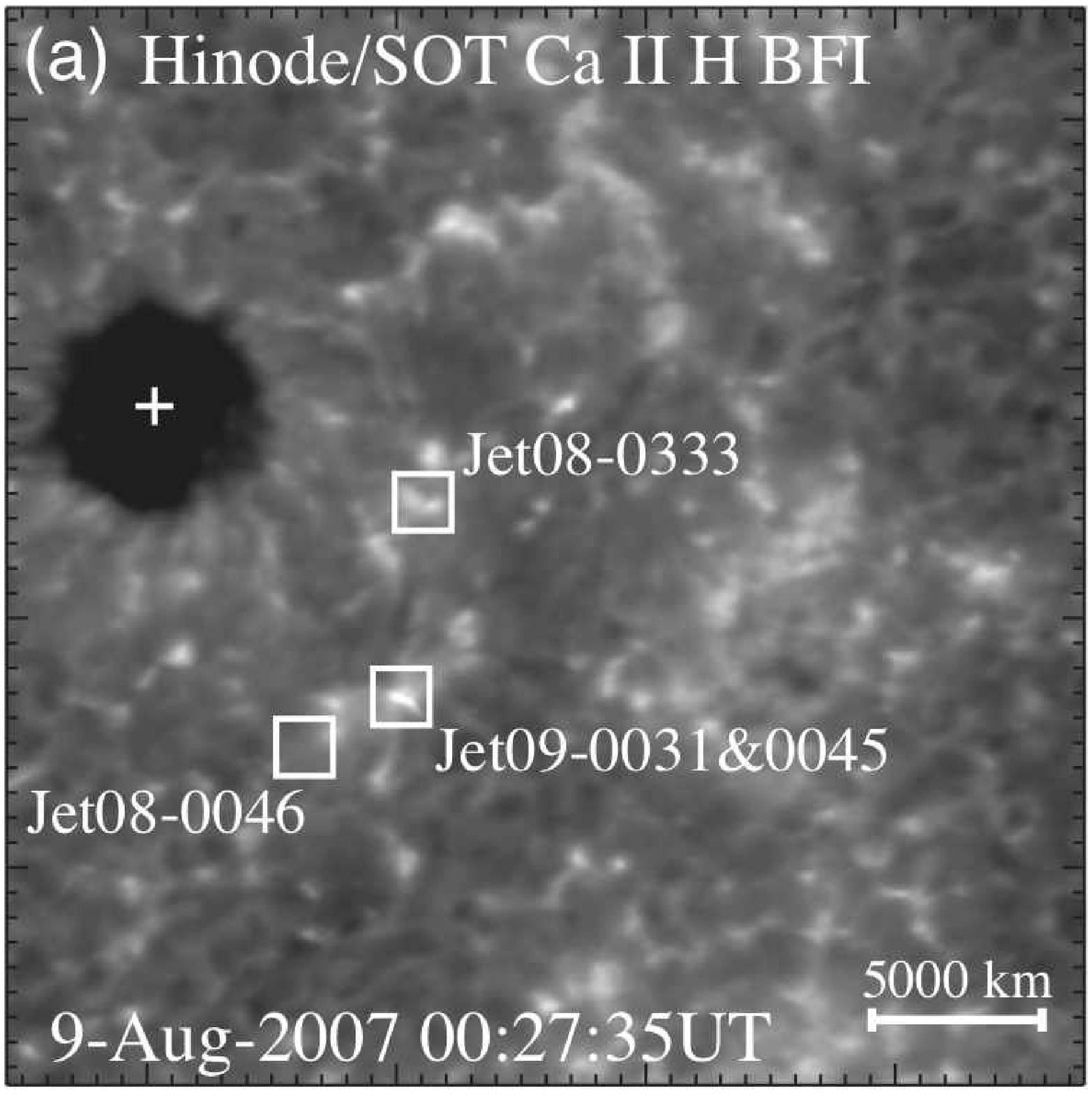}
    \FigureFile(78mm,78mm){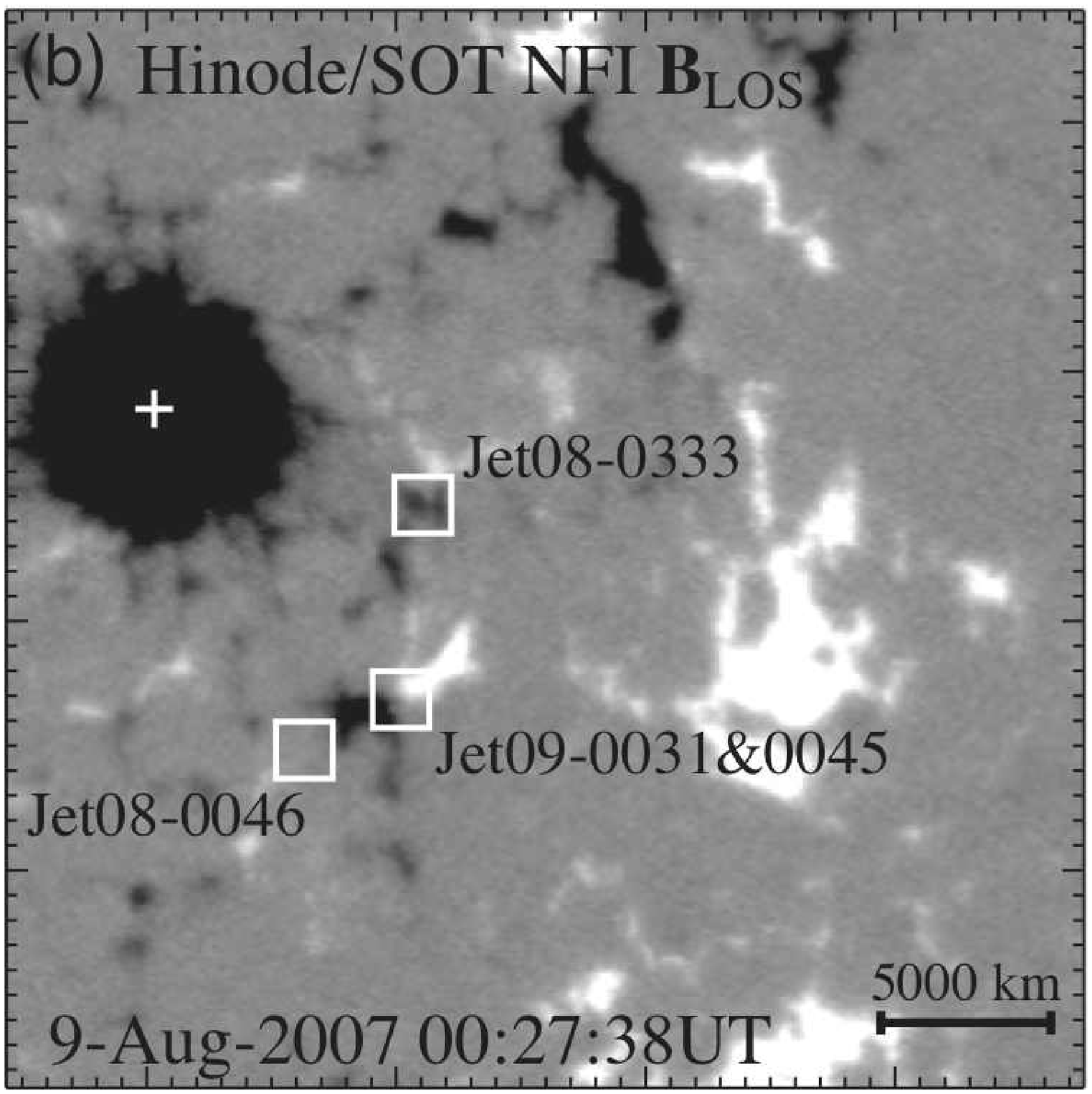}
  \end{center}
  \caption{The locations of four typical chromospheric anemone jets
    and their surrounding area (moat region of NOAA AR 10966) observed
    with (a) the Ca\emissiontype{II} H broadband filter equipped by
    the SOT/Hinode, and with (b) the Fe\emissiontype{I}
    6302\AA\ narrowband filter equipped by the SOT (Stokes V/I
    magnetogram): The time of the images are chosen for
    ``Jet09-0031''.  The boxes indicates the locations of the four
    typical jets, which we introduce in section~3.2.  Each jet
    location is measured on the local frame centered on the sunspot,
    and the location is transferred to the observed frame with the
    time of the images.  Top is north and left is east.  FOV of each
    image is 43.\arcsec2$\times$43.\arcsec2.}\label{fig:fov}
\end{figure}

The observations with the SOT/Hinode BFI were performed in 2$\times$2
summing mode (0\arcsec.109/pixel), covering an area of
56\arcsec$\times$112\arcsec\ (before 06:00 UT on August 8, 2007) or an
area of 83\arcsec$\times$83\arcsec\ (rest of the period) in the data
set.  The time cadence of the Ca\emissiontype{II} H BFI images was
15~seconds till 11:00 UT on August 7, 2007 and 40~seconds for the rest
of the period.  The normal calibration processes were performed on the
BFI images.  Thereafter the images were coaligned with each other
using the sunspot.

In order to investigate the magnetic environment surrounding the jets
as well as dynamical changes in the photospheric magnetic field at the
``footpoints of the jets'', we examined the Stokes data as well.  We
also used the SOT Narrowband Filter Imager (NFI) of FG observations
with the Stokes IQUV shutterless mode (cf. \cite{ichimoto2008}) for
the spectral band of Fe\emissiontype{I} 6302 \AA\ absorption line in
the HOP12.  The cadence of NFI Stokes observations were the same as
that for the BFI Ca\emissiontype{II} H images.  The time difference
between each NFI and BFI observations was 6 seconds.  The shutterless
mode Stokes IQUV data were available from 12:00 UT on August 7, 2007.
The data covered an area of 51\arcsec$\times$164\arcsec\ in 2$\times$2
summing mode (0\arcsec.16/pixel).  The passband of the narrowband
filter is 84m\AA\ (FWHM), that was positioned at -120m\AA\ of the
Fe\emissiontype{I} 6302.5 \AA\ line.  Dark subtraction, the correction
of bad pixels, and cosmic-ray removal were applied.  The polarization
calibration was applied to the NFI data following
\citet{ichimoto2008}.  Depending on the spectral band, scaling
corrections to BFI data and alignments within the NFI images and the
NFI and BFI images were applied.  A weak field approximation is
adopted for filtergrams (cf. \cite{landi2004}) to interpret the NFI
Stokes IQUV data for the construction of magnetograms.

Figure~\ref{fig:fov}a shows the Ca\emissiontype{II} H broadband filter
snapshot image of the AR 10966 at 00:27:35 UT on August 9, 2007, near
the disk center.  A typical chromospheric anemone jet (``Jet09-0031'')
is visible in the south-west region of the sunspot.  The length and
the width of the jet are 1000 km and 200 km, respectively.  The
velocity of the jet is estimated to be 9.7 km/s (see Table~1).  These
parameter values are similar to those of the chromospheric anemone
jet, reported by \citet{shibata2007} observed near the limb.

\subsection{Hida DST observation}

\begin{figure*}
  \begin{center}
    \FigureFile(170mm,120mm){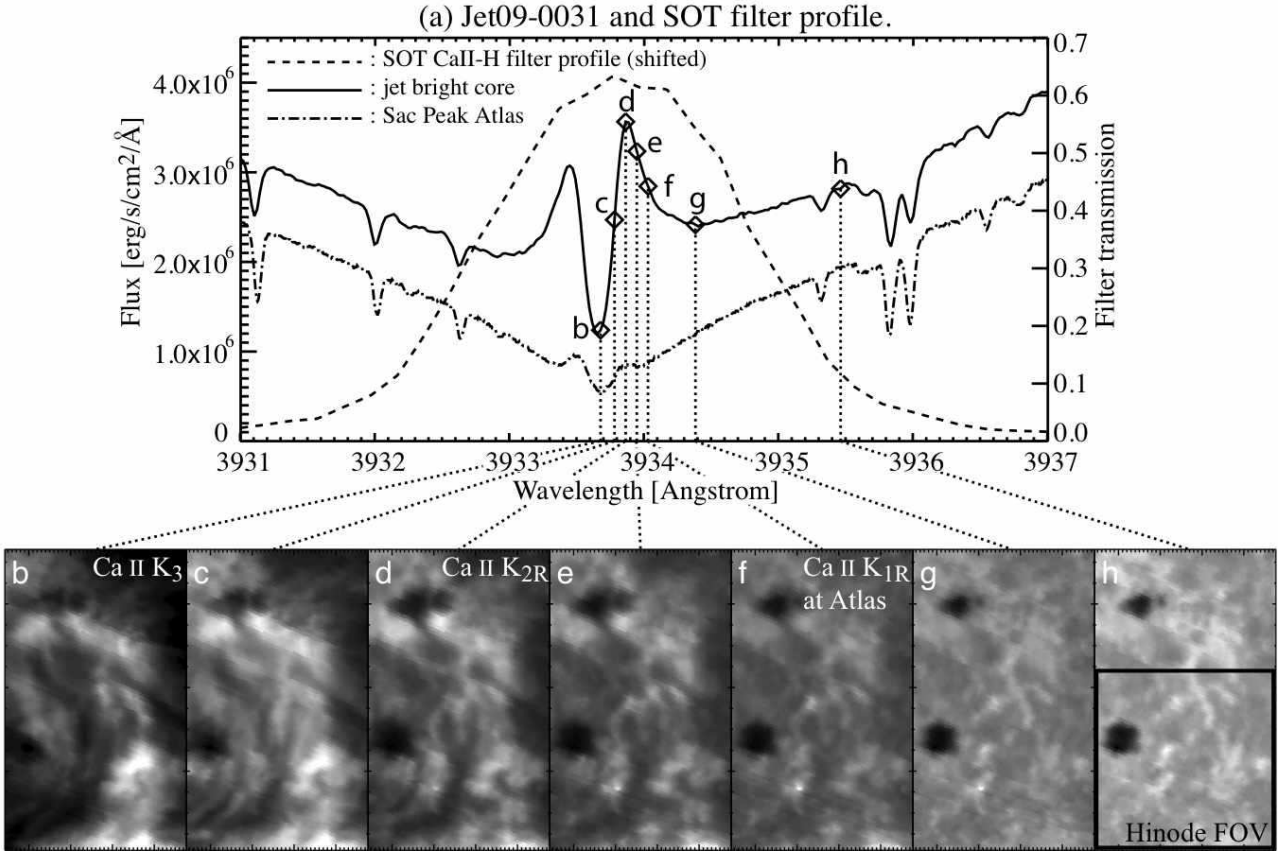}
  \end{center}
  \caption{(a) A typical Ca\emissiontype{II}~K line profile of a
    chromospheric anemone jet's bright core taken with the DST/Hida
    vertical spectrometer at 00:27:45 UT around the jet bright core in
    Figure 1: The solid curve in (a) represents the
    Ca\emissiontype{II}~K line profile of the jet bright core.  The
    broken curve in (a) shows the SOT/Hinode Ca\emissiontype{II}~H
    filter profile, which is shifted to the Ca\emissiontype{II}~K
    line.  The dash-dotted curve in (a) shows the
    Ca\emissiontype{II}~K line profile from the Sac Peak Atlas.
    Frames (b)--(h) show the spectroheliograms at the wavelengths
    indicated by diamonds in (a).  Top is north and left is east.
    Each image in (b)--(h) has the same FOV of
    0.72\arcmin$\times$1.2\arcmin.  The smaller FOV in (h) represent
    the SOT/Hinode FOV of Fig.~\ref{fig:fov}.}\label{fig:diagram}
\end{figure*}

The observations for the Ca\emissiontype{II} K spectroheliograph in
this study were obtained with the vertical spectrograph of the 60~cm
Domeless Solar Telescope of the Hida Observatory, Kyoto University.
The wavelength coverage was about 16 \AA\ around the core of
Ca\emissiontype{II} K line (3933.68 \AA).  The wavelength sampling was
0.021 \AA\ per pixel.  The spectrograph slit had a width of 50~$\mu$m
(0\arcsec.32 on the sun), and scanning step and range were 0\arcsec.40
and 98\arcsec, respectively.  The slit length corresponded to about
144\arcsec\ on the sun, with the pixel resolution along the slit of
0\arcsec.24 per pixel.  Dark subtractions and flat-field corrections
were applied to each of the spectrograms in the sequence.  The
wavelengths of two photospheric lines (Fe\emissiontype{I} blend
3932.64 \AA\ and Fe\emissiontype{I} 3935.32 \AA), each averaged along
the entire length of the slit, were used as the wavelength references
for each exposure.  Intensity corrections to the spectroheliograms
were applied by fitting the averaged quiet region spectrum with the
Sacramento Peak Atlas \citep{beckers1976} so that the intensity unit
of a spectroheliogram becomes erg/s/cm$^{2}$/\AA.  Thereafter the
observed intensity is converted into the flux emitted at the solar
surface with assuming isotropic radiation from jets and their
surrounding area.  The distance between the sun and the Earth
($1.52\times10^{8}$ km) and the solar radius ($6.96\times10^{5}$ km)
are used for this conversion.

Figure~\ref{fig:diagram} shows a Ca\emissiontype{II} K line profile of
a chromospheric anemone jet (``Jet09-0031'') taken with the DST/Hida
vertical spectroheliograph.  The broken curve in
Fig.~\ref{fig:diagram}a shows the SOT/Hinode Ca\emissiontype{II} H
filter profile, which is shifted to the Ca\emissiontype{II} K line.
Figures~\ref{fig:diagram}b--\ref{fig:diagram}h show spectroheliograms
made in various wavelengths shown in Fig.~\ref{fig:diagram}a.

DST/Hida also equips a 0.25 \AA\ passband H$\alpha$ Lyot filter, and
takes images at five wavelength positions (H$\alpha$ center and its
wing at $\pm$0.5\AA\ and $\pm$0.8\AA) without disturbing the
Ca\emissiontype{II} K spectroheliograph observation.  We use these
H$\alpha$ Lyot filter images for extracting additional spectral
information in the chromosphere, and for understanding the magnetic
connectivities in the chromosphere in the region around jets.  The
details of H$\alpha$ observations are discussed in the Appendix.

\section{Results}

\subsection{Overview of the observed region}

\begin{table*}
  \begin{center}
  \caption{Basic Data of SOT Ca\emissiontype{II} Jets around NOAA AR10966.}\label{tbl_jets}
    \begin{tabular}{crcrcccrrc}
      \hline 
      Event & SOT Ca\emissiontype{II} H & $\mu$\footnotemark[$*$] & Life  & Max I/ & Area          & Max    
      & V$_{\perp}$ & V$_{LOS}$\footnotemark[$\ddagger$] & Ca\emissiontype{II} K$_{1}$/K$_{2}$\\
      Name   &  Peak time    &            & time  & quiet I & size\footnotemark[$\dagger$] & length & 
      												   &  &  asymmetry\\ 
               &   (UT)            &            &     (sec)    &            & (Mm$^2$) & (Mm)  & (km/s) & (km/s) & \\
      \hline
Jet07-0301 &    2007 Aug 07 03:01 & 0.82 & 180 & 1.79 & 1.23 & 1.16 &  6.5 & -5.7 & red \\
Jet08-0046 &         Aug 08 00:46 & 0.92 & 320 & 1.95 & 1.85 & 1.49 &  4.6 & -0.8 & blue \\
Jet08-0319 &                03:19 & 0.93 &  40 & 1.79 & 1.99 & 2.81 &  2.8 &  0.0 & red \\
Jet08-0333 &                03:33 & 0.93 & 170 & 1.53 & 1.85 & 2.71 & 13.6 & -1.9 & red \\
Jet08-0430 &                04:30 & 0.93 & 200 & 1.61 & 0.65 & 4.81 & 24.1 &  2.7 & not obvious\\
Jet09-0031 &         Aug 09 00:31 & 0.97 & 270 & 2.02 & 0.49 & 0.86 &  9.7 & -2.2 & red \\
Jet09-0045 &                00:45 & 0.97 & 240 & 2.28 & 0.73 & 0.45 &  5.1 & -2.3 & red \\
Jet09-2350 &                23:50 & 0.97 & 180 & 1.85 & 0.37 & 0.56 &  3.1 & -3.4 & red \\
Jet10-0011 &	     Aug 10 00:11 & 0.97 &  90 & 2.06 & 0.33 & 1.02 & 11.2 &  0.0 & not obvious\\
      \hline
      \\
      \multicolumn{9}{@{}l@{}}{\hbox to 0pt{\parbox{180mm}{\footnotesize
	\par\noindent
	\footnotemark[$*$] $\mu = cos\theta$, where $\theta$ is the heliocentric angle.
	\par\noindent
	\footnotemark[$\dagger$] area size of bright features at SOT Ca\emissiontype{II} H intensity peak time.
	\par\noindent
	\footnotemark[$\ddagger$] by Doppler shift of the Ca\emissiontype{II} K$_3$ absorption line. ``blue shift'' is negative.
        Estimated error is $\pm1.6$ km/s. 
	}\hss}}
    \end{tabular}
  \end{center}
\end{table*}

The active region AR 10966 had an umbra with negative magnetic
polarity, while the southwestern half of the moat region was dominated
by positive magnetic polarity (see Fig.~\ref{fig:fov}b).  A boundary
of the magnetic polarities in the moat region and the sunspot was
located in the west side of the umbra, and this is a favorable site
for the emerging flux in the active region.  A strong activity of
emerging flux started in this part of the active region around 22 UT
on August 7, 2007, then a new umbra was formed in the north side of
the original umbra around 12 UT on August 8, 2007 (see
Fig.~\ref{fig:diagram}).

During the coordinated observation, 9 chromospheric anemone jets were
observed simultaneously with Hinode Ca\emissiontype{II} H broadband
filter and DST/Hida Ca\emissiontype{II} K spectroheliograms.  Table~1
shows the basic properties of 9 jets.  In the following section,
evolution and dynamics of four typical jets out of these 9 jets are
discussed.  All of the four jets occurred with magnetic flux
cancellations at the photosphere.  The opposite polarity magnetic
elements around the cancellation sites were converged by the
extensions of EFRs or Moat flow.

\

\subsection{Case study}

\subsubsection{Jets associated with emerging flux (Aug. 9 0031 and 0045)}

\begin{figure}
  \begin{center}
    \FigureFile(80mm,60mm){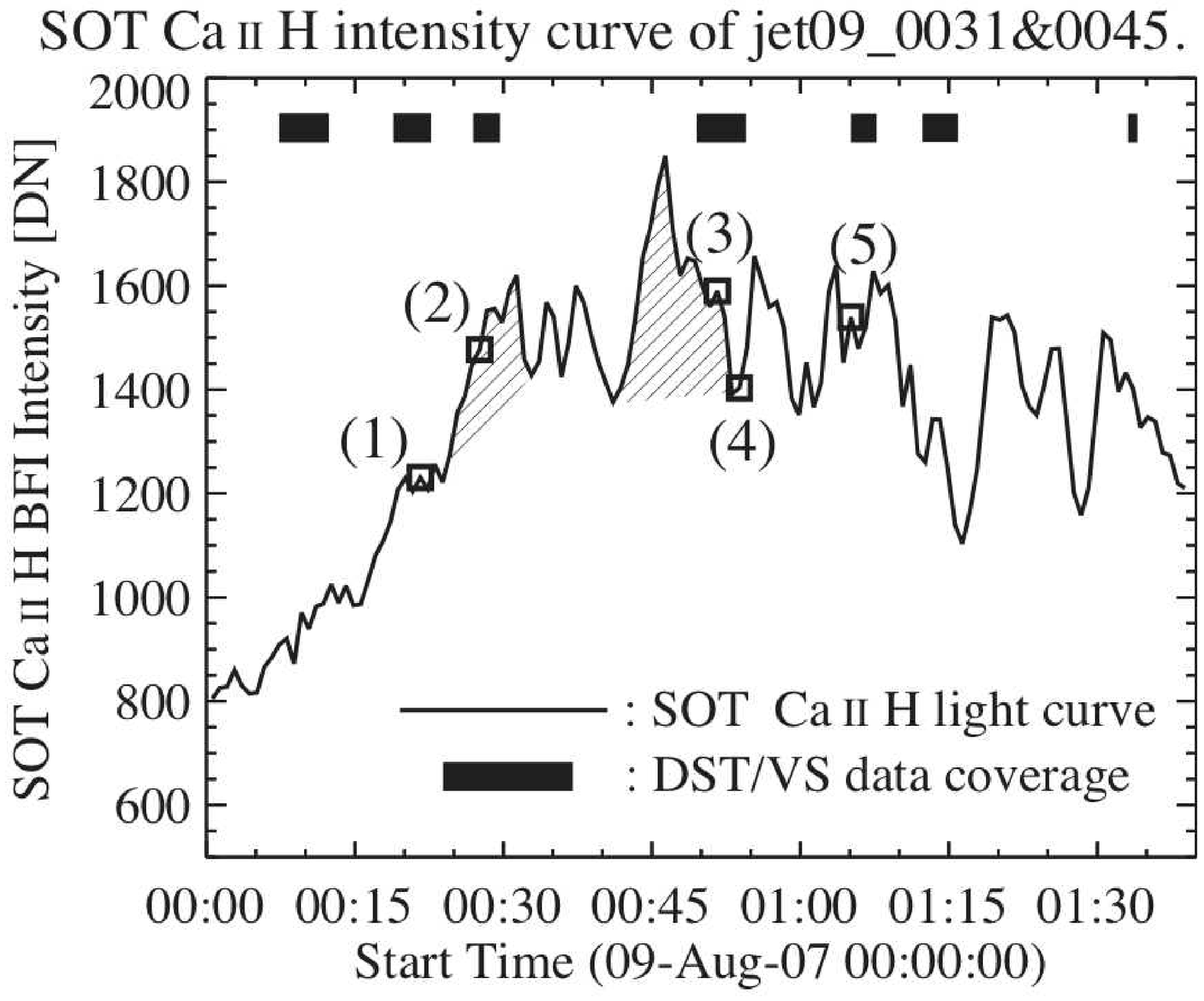}
  \end{center}
  \caption{Time series of the SOT BFI Ca\emissiontype{II}~H intensity
    curve (solid line) at the locations on the same polarity inversion
    line where the ``Jet09-0031'' occurred: This intensity curve was
    made as tracking the locations of the same polarity inversion
    line, which was moving (cf. Fig.~\ref{fig:evoljet1}). Sampled area
    is 1.\arcsec09 $\phi$ ($\sim$800km $\phi$) disk, and the
    intensities are averaged in it (in arbitrary unit).  The first
    large peak represents the ``Jet09-0031'', and the strongest peak
    around 00:45 UT represents the ``Jet09-0045'' (see the peaks
    indicated by oblique hatching). Each square with a number on this
    curve corresponds to the timing of a row in
    Fig.~\ref{fig:evoljet1} and of a Ca\emissiontype{II}~K spectra in
    Fig.~\ref{fig:evoljet1spec}, with the corresponding number,
    respectively.  The horizontal black bars near the top of this
    figure show the DST/VS data coverage.}\label{fig:jet1lcurve}
\end{figure}

\begin{figure*}
  \begin{center}
    \FigureFile(170mm,138mm){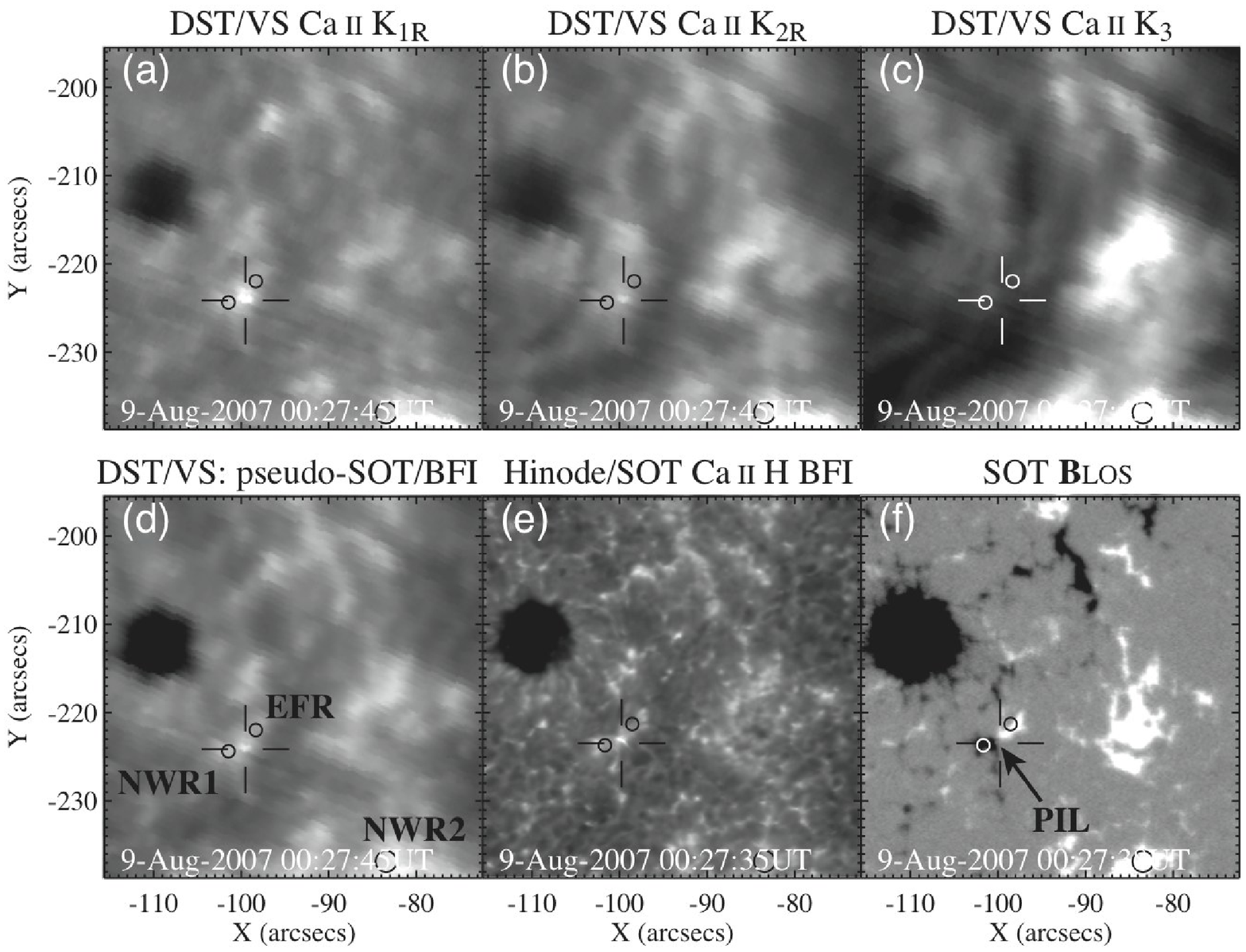}
   \end{center}
  \caption{The DST/Hida Ca\emissiontype{II}~K spectroheliograms of (a)
    Ca\emissiontype{II}~K$_{1R}$, (b) Ca\emissiontype{II}~K$_{2R}$,
    (c) Ca\emissiontype{II}~K$_3$, and (d) a pseudo
    Ca\emissiontype{II}~K broadband image made of the DST
    Ca\emissiontype{II}~K spectroheliograms and the SOT
    Ca\emissiontype{II}~H broadband filter profile (cf.
    Fig.~\ref{fig:diagram}a), and the corresponding SOT images of (e)
    BFI Ca\emissiontype{II}~H, and (f) NFI photospheric magnetogram:
    Top is north and left is west.  Each image has the same FOV of
    Fig.~\ref{fig:fov}.  The cross-hairs indicates the jet core
    brightening location.  The circles indicate the locations of an
    EFR and network regions (EFR, NWR1, and NWR2), in which the
    average spectra are shown in Fig.~\ref{fig:profiles1}.  White and
    black in frame (f) represent positive and negative magnetic
    polarities (color is saturated at $\pm$300 Gauss).  The arrow
    indicates the polarity inversion line (PIL) where the jet
    occurred. }\label{fig:dstbfinfi}
\end{figure*}

It is found that the location of the jets lies near the boundary
between the positive and negative magnetic polarities or the polarity
inversion line (PIL in Fig.~\ref{fig:dstbfinfi}f).  A series of jets
(five or more) has occurred on the same polarity inversion line during
a period of around 45 minutes, centered in around 00:45 UT on August
9, 2007.  Figure~\ref{fig:jet1lcurve} shows the time variation of the
BFI Ca\emissiontype{II} H intensity at the locations on the same
polarity inversion line (see PIL in Fig.~\ref{fig:dstbfinfi}f).  Every
single intensity peak in the light curve represents a ``single'' jet
event.

The first large peak indicated by oblique hatching in
Fig.~\ref{fig:jet1lcurve} represents the ``Jet09-0031'' (see Table~1),
which was seen in Fig.~\ref{fig:fov} and ~\ref{fig:diagram}.  The
spectroheliograms in Ca\emissiontype{II} K$_{1R}$, K$_{2R}$, and K$_3$
of this jet event at 00:27:45UT (3 minutes before the
Ca\emissiontype{II} H intensity peak; see Fig.~\ref{fig:jet1lcurve})
along with its surrounding area are shown in
Figures~\ref{fig:dstbfinfi}a--c.  Fig.~\ref{fig:dstbfinfi}d shows the
pseudo-Hinode Ca\emissiontype{II} H image for this region, and
Fig.~\ref{fig:dstbfinfi}e depicts the SOT BFI Ca\emissiontype{II}~H
image from Fig.~\ref{fig:fov} for comparison.  Although the spatial
resolution is very different (0.\arcsec2 for Hinode and 1\arcsec\ for
DST/Hida with seeing condition), overall morphology is still similar.
Comparison between the four Ca\emissiontype{II} spectroheliograms and
Hinode Ca\emissiontype{II}~H image reveals that the Hinode
Ca\emissiontype{II} jet is formed below the middle chromosphere, since
the jet cannot be seen in K$_3$ image.  Fig.~\ref{fig:dstbfinfi}f
shows the SOT/Hinode NFI photospheric magnetogram of the same field of
view.  It is found the polarity inversion line was formed by the
continuous collision between a positive polarity source of a newly
emerging dipole (shown by black circle in Fig.~\ref{fig:dstbfinfi}f;
EFR), and an isolated negative source in the moat region (open circle
in Fig.~\ref{fig:dstbfinfi}f; NWR1).

\begin{figure}
  \begin{center}
    \FigureFile(80mm,61mm){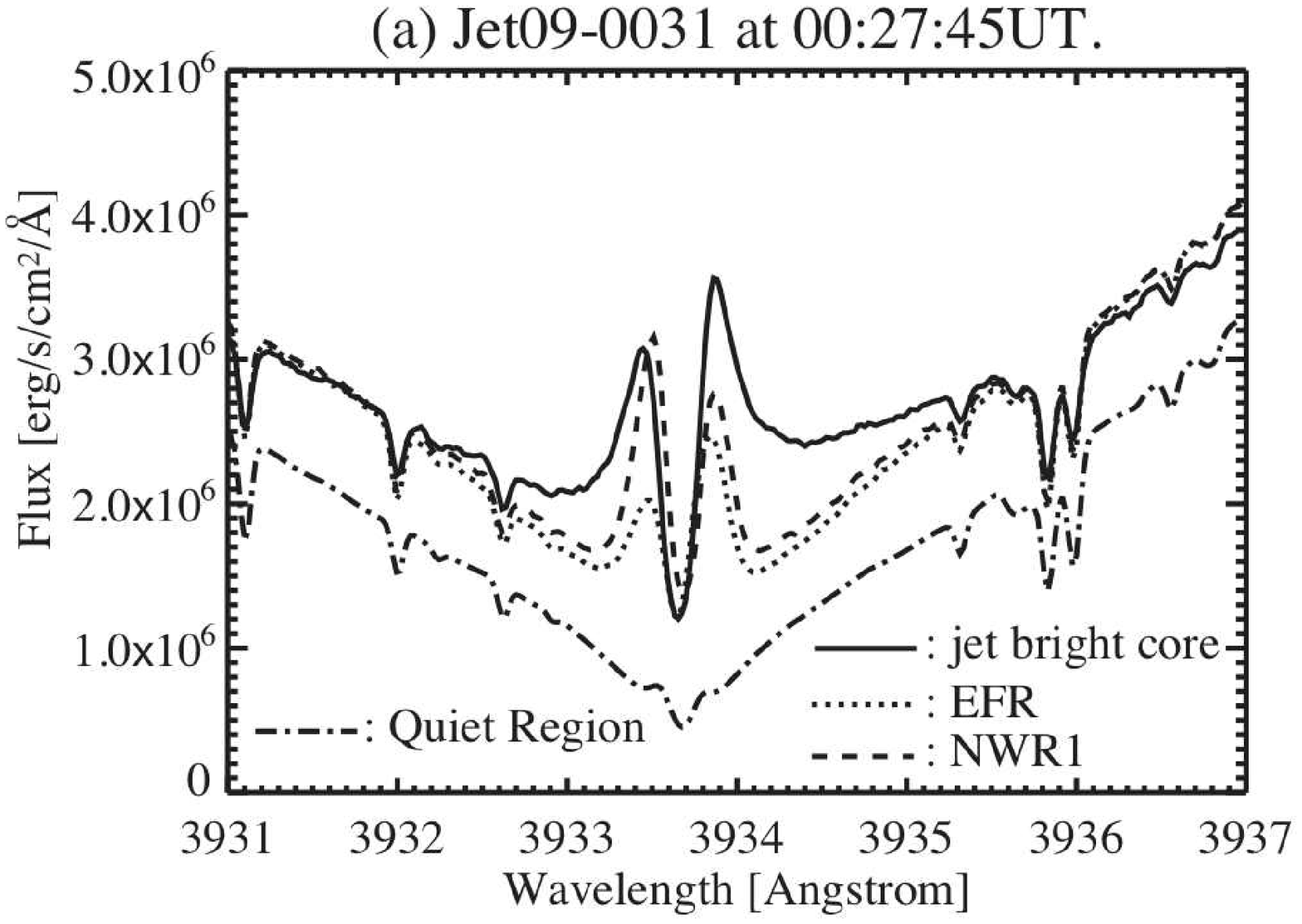}
    \FigureFile(80mm,61mm){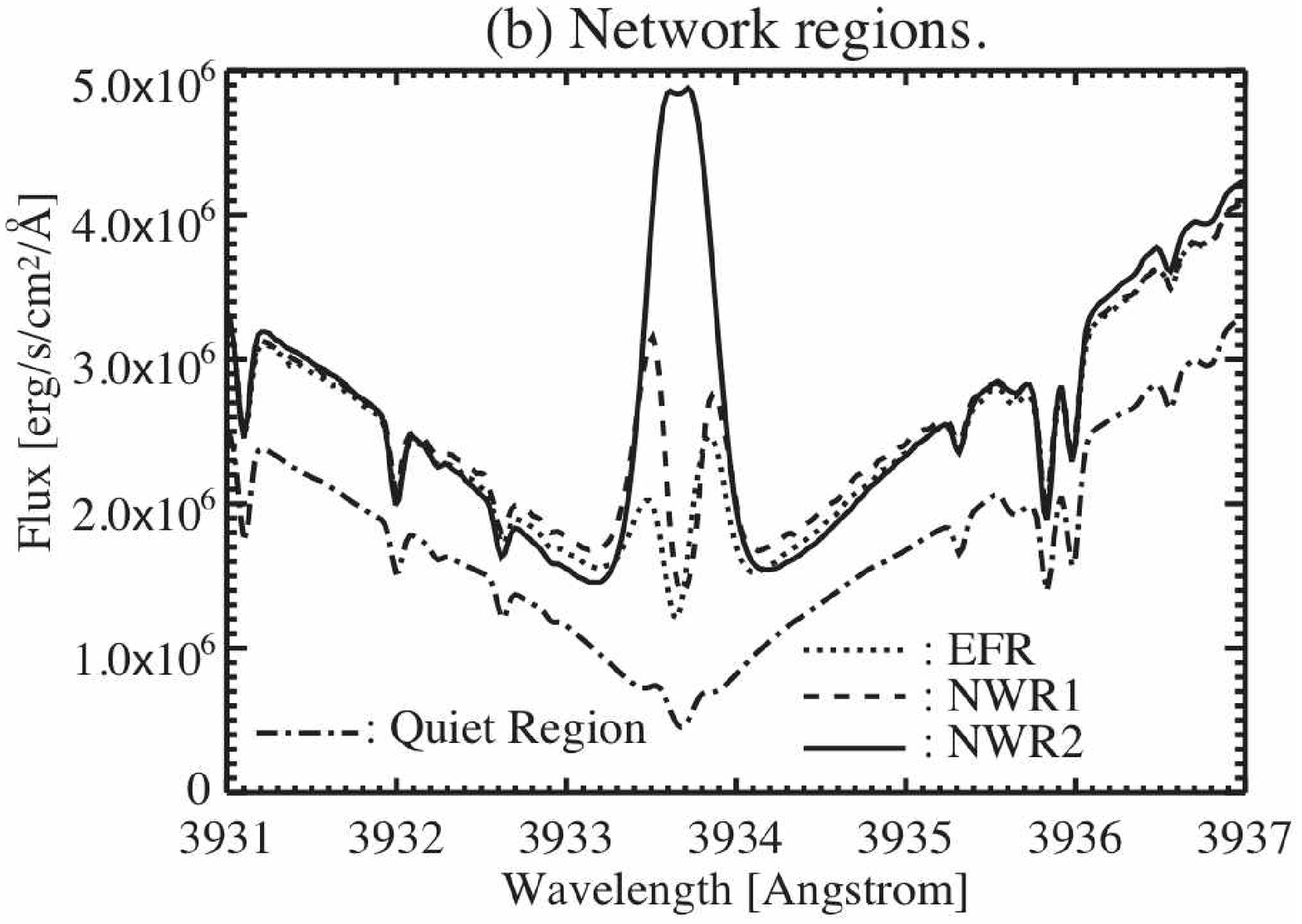}
    \FigureFile(80mm,60mm){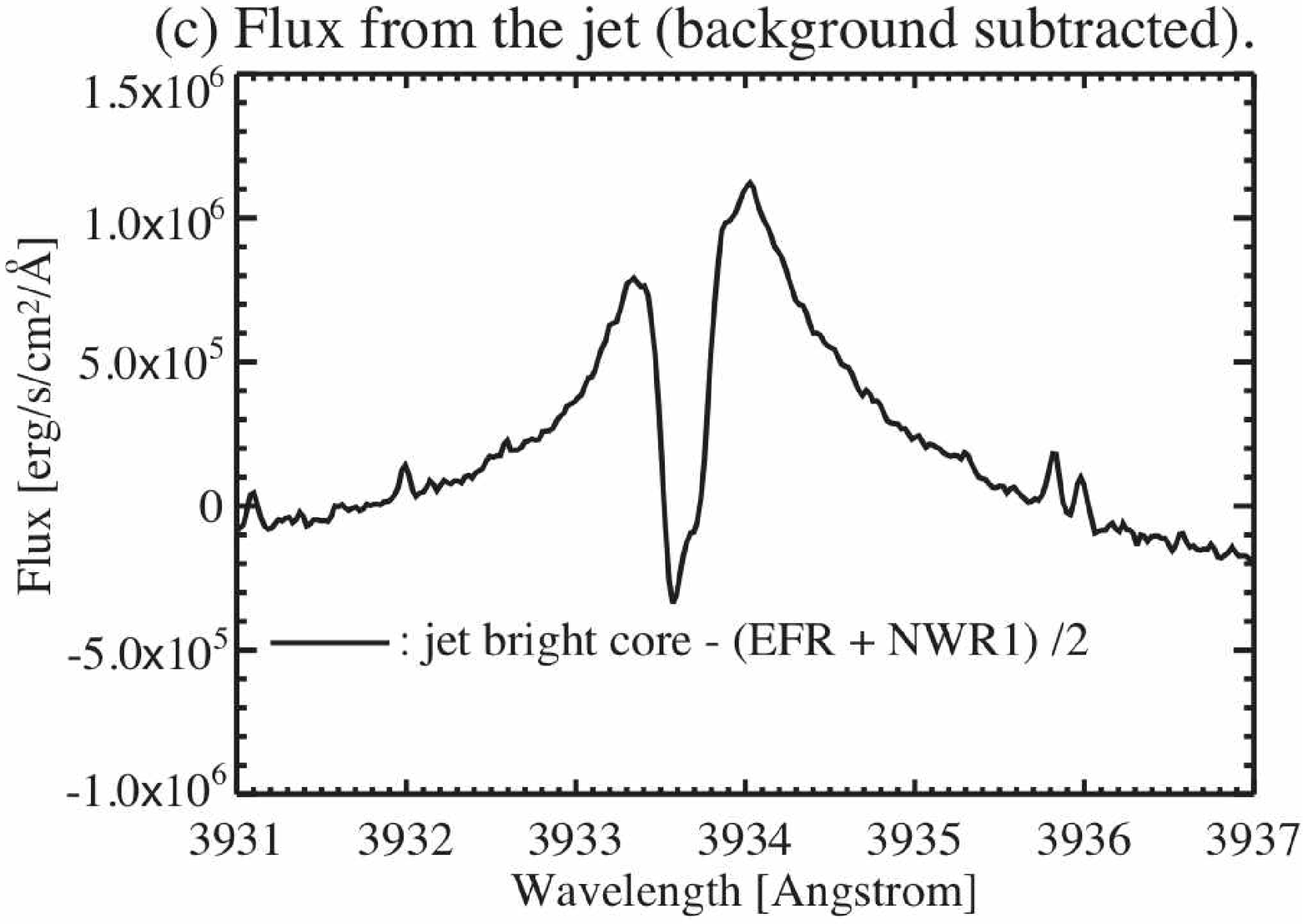}
  \end{center}
  \caption{Ca\emissiontype{II} K spectra of the ``Jet09-0031'' at
    00:27:45UT of (a) jet bright core at the center of the cross-hairs
    in Fig.~\ref{fig:dstbfinfi} (solid curve), an EFR and a network
    region near by this jet (EFR and NWR1; dotted and broken curves),
    and quiet region outside of the moat region (dash-dotted curve),
    (b) EFR and network regions at two locations (EFR, NWR1, and NWR2)
    indicated in Fig.~\ref{fig:dstbfinfi}, and (c) subtracted profile
    of (``jet bright core'' $-$ ``(EFR$+$NWR1)/2''): The vertical axis
    shows the flux emitted at the solar surface.  Each spectrum is
    made with averaging the spectra in a sample
    region.}\label{fig:profiles1}
\end{figure}

Figure~\ref{fig:profiles1} shows the Ca\emissiontype{II}~K spectra of
the ``Jet09-0031'' at the time of the spectroheliograms in
Fig.~\ref{fig:dstbfinfi}.  The solid curve in
Fig.~\ref{fig:profiles1}a represents the averaged spectrum at the jet
bright core.  In order to demonstrate the emission increased by the
jet, we compared the Ca\emissiontype{II}~K emission spectrum from the
jet bright core with the spectra from the nearest
Ca\emissiontype{II}~K network regions of this jet (dotted and broken
curves).  The spectrum from the jet bright core shows a characteristic
intensity increase from the spectra of the nearest network regions,
widely ($\sim4.8$\AA) around K$_{1}$.  We noticed that this intensity
increase around K$_{1}$ is possibly due to the presence of jet.  The
spectra from the nearest Ca\emissiontype{II} network regions show
similar intensity variation with spectra of the Ca\emissiontype{II}
network regions in the field of view, and larger than that of the
quiet region (Fig.~\ref{fig:profiles1}b).  The difference in the
K$_{3}$ core of the network regions' spectra, as shown in
Fig.~\ref{fig:profiles1}b, comes due to the existence of dark filament
above the jet region (see Fig.~\ref{fig:dstbfinfi}c and H$\alpha$
images in Fig.~\ref{fig:dstha1} for more details).
Fig.~\ref{fig:profiles1}c shows the intensity of the jet subtracted by
the background network intensity.  This intensity distribution will be
used later (in section~4) to estimate the released energy during the
jet event.

\begin{figure*}[p]
  \begin{center}
    \FigureFile(150mm,180mm){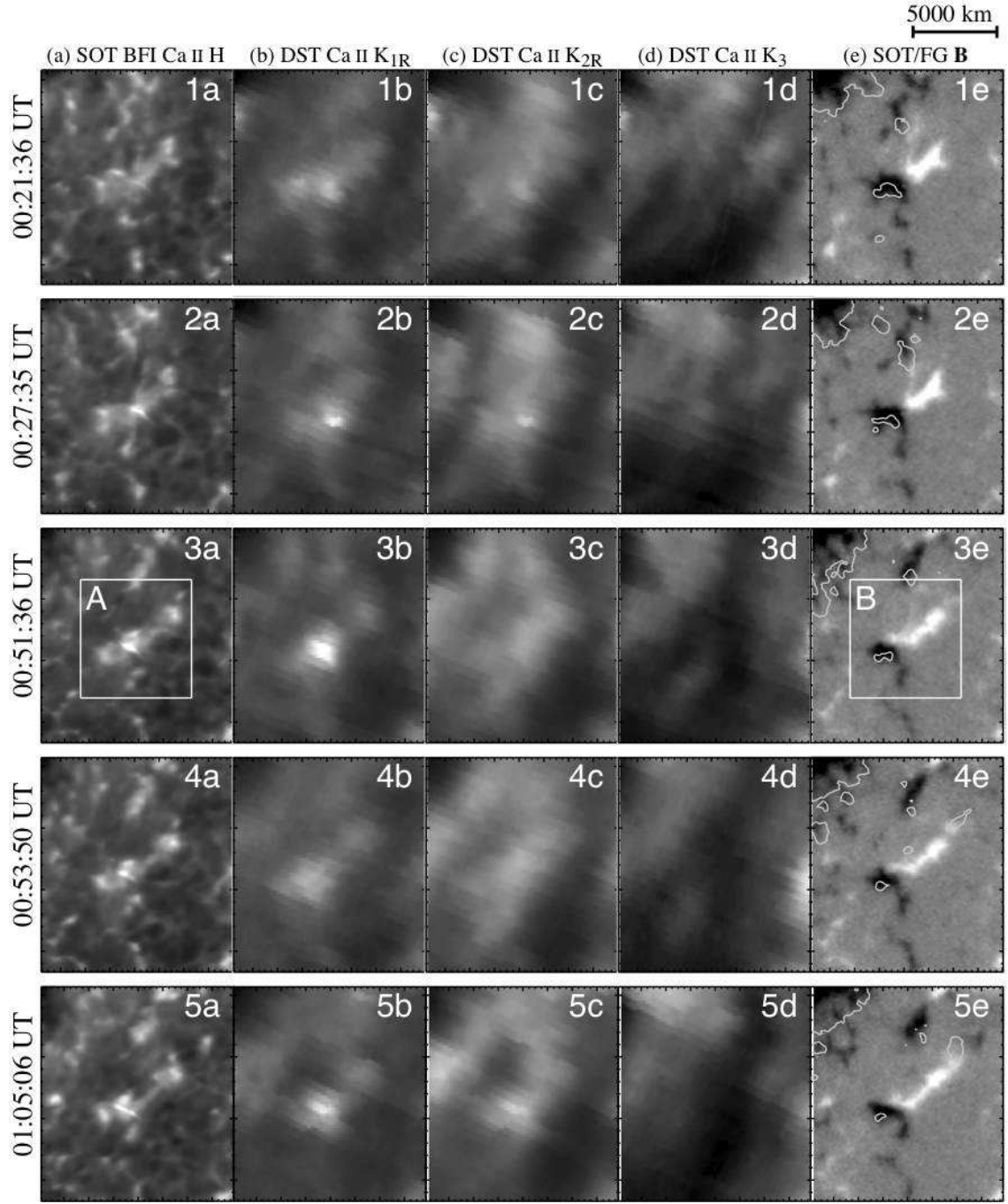}
  \end{center}
  \caption{Evolutions of a series of chromospheric jet events on
    August 9 around 00:45 UT: (a) SOT Ca\emissiontype{II}~H BFI
    images.  DST/Hida spectroheliograms of (b)
    Ca\emissiontype{II}~K$_{1R}$, (c) Ca\emissiontype{II}~K$_{2R}$,
    and (d) Ca\emissiontype{II}~K$_3$. (e) SOT NFI Fe\emissiontype{I}
    6302~\AA\ magnetograms. The time of rows are chosen as correspond
    to the squares with the same numbering in
    Fig.~\ref{fig:jet1lcurve}.  Top is north and left is east.  Each
    image has the same FOV of 15.\arcsec6$\times$14.\arcsec4. In
    frames~(e), white and black colors represent positive and negative
    line-of-sight component of the photospheric magnetic field {\bf B}
    (color is saturated at $\pm$300 Gauss), while the contours show
    the horizontal components of {\bf B} (contour level: 200 Gauss).
    The smaller FOVs ``A'' and ``B'' in the 3rd row represent the FOV
    of Fig.~\ref{fig:evoljet1detailed}.  In this figure, three
    chromospheric jets occurred on the same polarity inversion line.
    The 2nd, 3rd, and 5th rows show ``Jet09-0031'', ``Jet09-0045'',
    and another minor jet, respectively.}\label{fig:evoljet1}
\end{figure*}

\begin{figure}
  \begin{center}
    \FigureFile(85mm,60mm){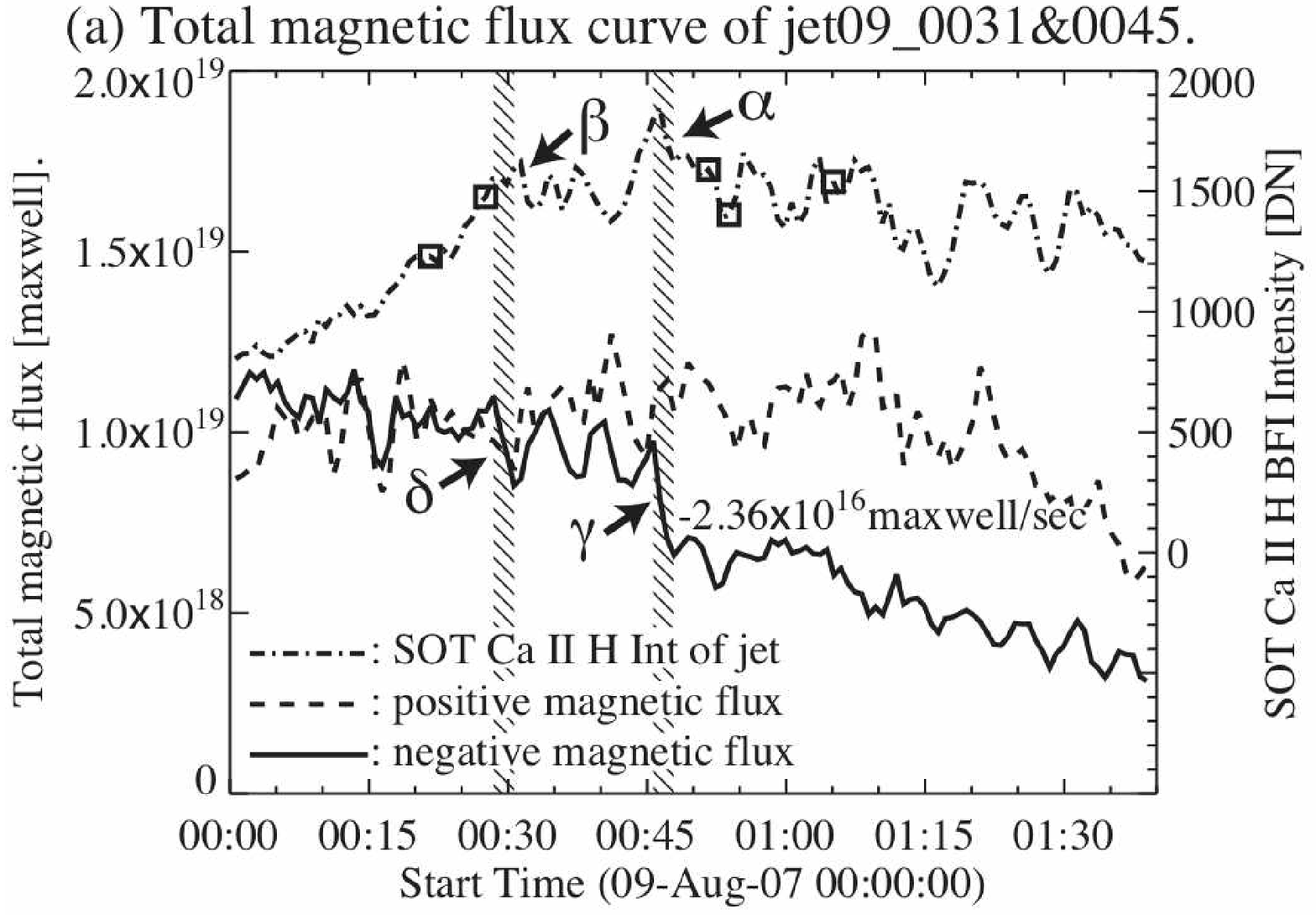}
    \FigureFile(85mm,60mm){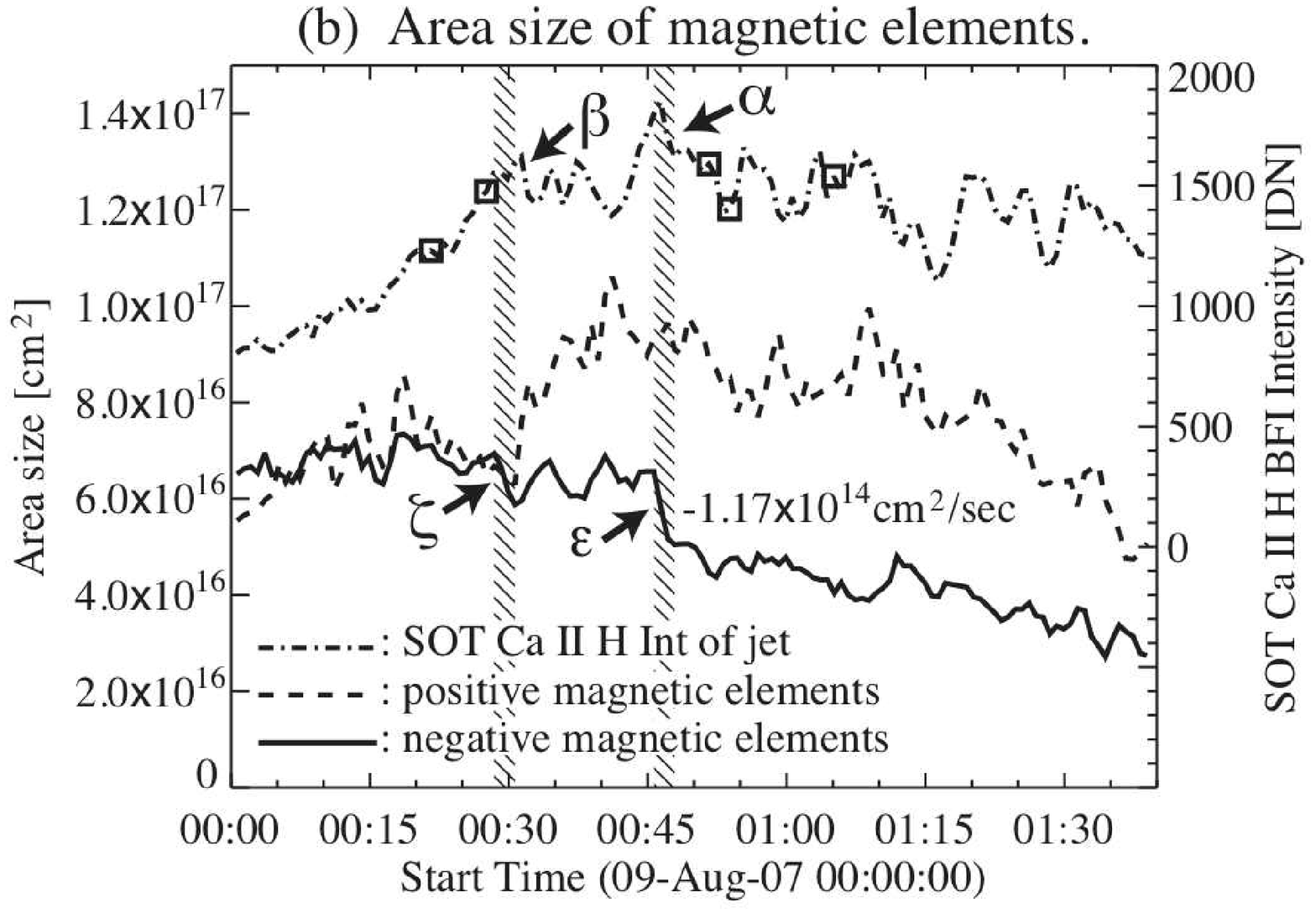}
  \end{center}
  \caption{Evolutions of (a) total magnetic flux and (b) area size of
    the magnetic sources that involved to the series of jet events on
    August~9 around 00:45 UT: The solid curves and the broken curves
    represent the plots for the isolated negative and positive
    magnetic sources in Fig.~\ref{fig:evoljet1detailed}, respectively.
    Samplings are made by using the line-of-sight component of the
    photospheric magnetic field with SOT/FG magnetograms.  Cut-off
    intensity for samplings is 42 Gauss in absolute value.  The
    Ca\emissiontype{II} intensity curve from Fig.~\ref{fig:jet1lcurve}
    is plotted as a reference of the timings of the jets (dash-dotted
    curves).  A distinct magnetic flux cancellation (slopes $\gamma$
    and $\epsilon$ on the curves for the negative polarity) is seen at
    the same timing of ``Jet09-0045'' (peak $\alpha$ on the
    Ca\emissiontype{II}~H light curves).  Another coincidence is also
    seen with ``Jet09-0031'' (see $\beta$, $\delta$, and $\zeta$).
    The oblique hatching indicates the timings of those magnetic flux
    cancellations.}\label{fig:jet1tflux}
\end{figure}

Figure~\ref{fig:evoljet1} shows snapshots of the evolutions of the
region above the PIL, with enlarge the area around the jets.  The jets
occurred intermittently.  We find that the timings and the shapes of
the emission features in K$_{1R}$ correlated well with the occurrences
of the jets in the Ca\emissiontype{II} H broadband filter images.  The
K$_{2R}$ counterpart of the jet was seen only during onset of
``Jet09-0031''.  No K$_{3}$ counterpart were seen in any frame, while
a K$_{3}$ filament was seen covering the jet region with the same
alignment of the array of the photospheric magnetic sources that are
involved in the series of the jet events.  It is found that all the
jets occurred along the same polarity inversion line.  The occurrence
of the ``magnetic flux cancellation'' around the polarity inversion
line is suggestive of shrinking of the isolated negative polarity.
This is confirmed by the curves of the total magnetic flux and the
area size of the negative magnetic sources shown in
Figure~\ref{fig:jet1tflux}.  The positive polarity source that formed
the polarity inversion line didn't show the ``shrinkage'' because this
magnetic source consists of a newly emerging dipole (EFR) and new flux
were supplied intermittently from the north.

\begin{figure}
  \begin{center}
    \FigureFile(80mm,145mm){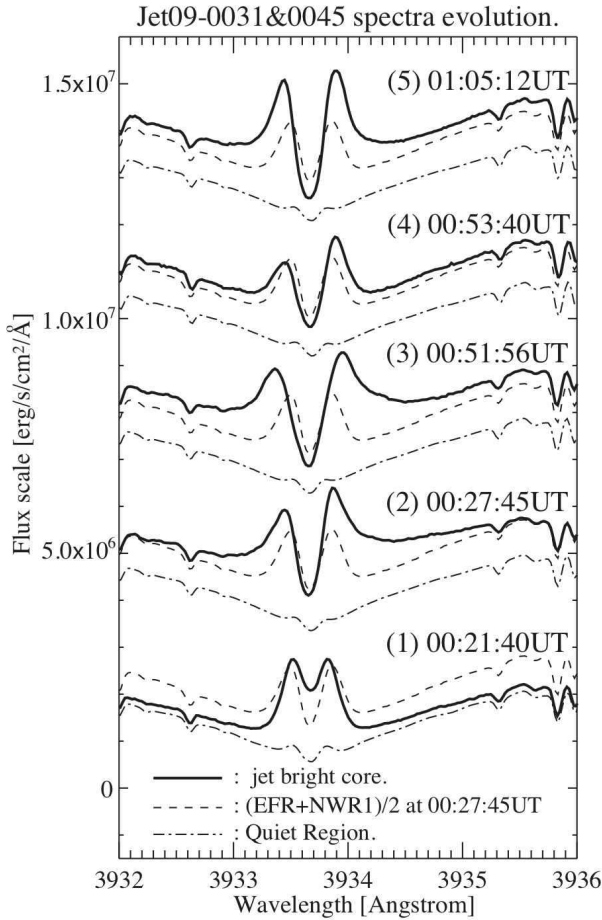}
  \end{center}
  \caption{Evolution of Ca\emissiontype{II}~K spectra of the jets'
    bright cores in the series of jet events on August~9 around 00:45
    UT: The time evolution in this figure is from the bottom to the
    top.  The time of each spectrum is chosen as the same timing of a
    row in Fig.~\ref{fig:evoljet1} with the corresponding number.  The
    solid curves indicate the spectra of the jet bright cores ((2),
    (3), and (5)), or those of the corresponding locations but rather
    in quiet condition ((1) and (4)).  Each spectrum is made with
    averaging the spectra in a sample region (0.\arcsec96 $\phi$
    ($\sim$700km $\phi$) disk).  Two reference spectra (of quiet
    region (dash-dotted curve) and of the network region near by the
    jet bright core at 00:27:45 UT (broken curve)) are over plotted
    for showing intensity increases with jets.  Basically, jet bright
    cores in this event series are ``red asymmetry'' and bright at
    K$_{1}$.}\label{fig:evoljet1spec}
\end{figure}

\begin{figure}
  \begin{center}
    \FigureFile(80mm,56mm){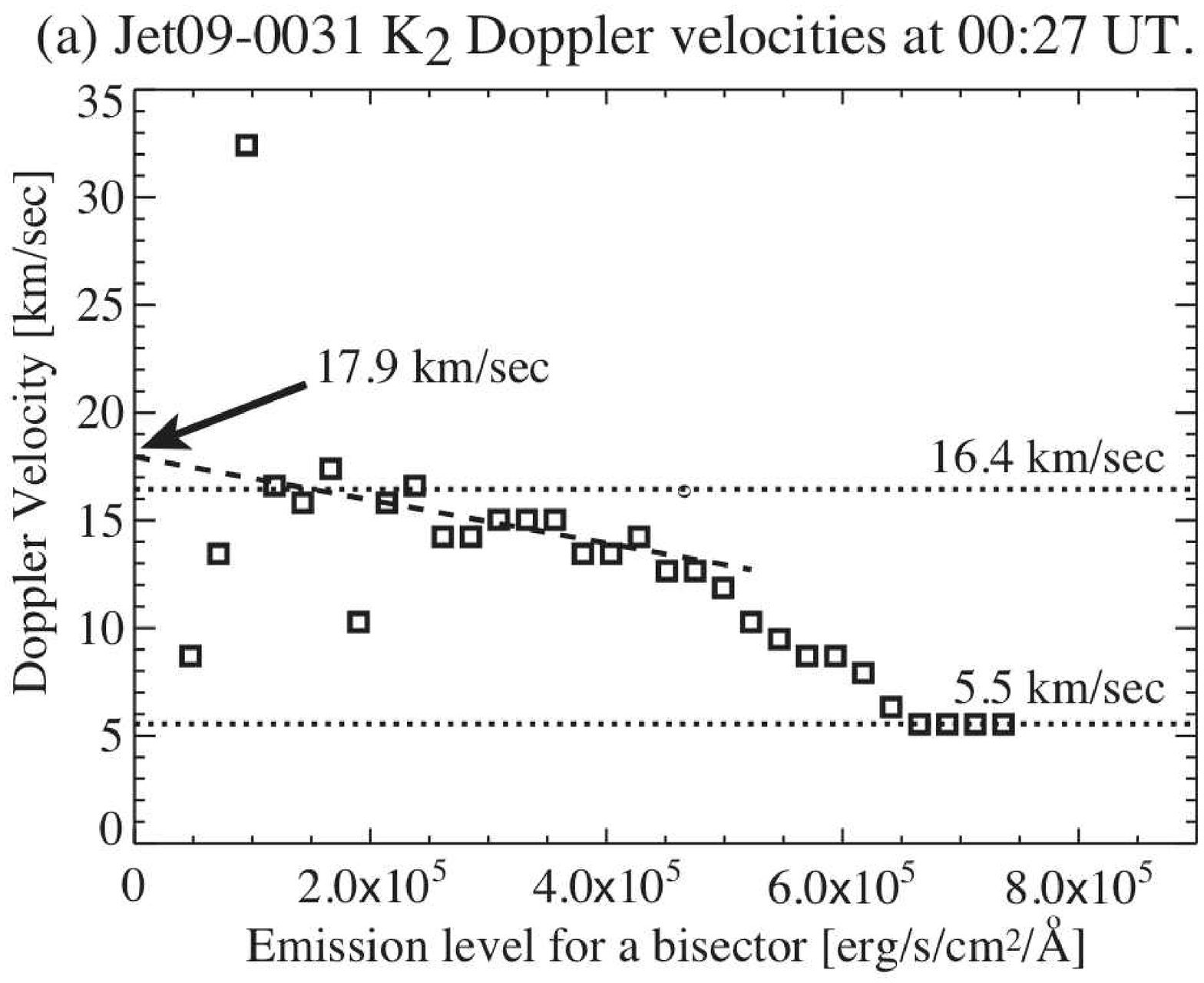}
    \FigureFile(80mm,56mm){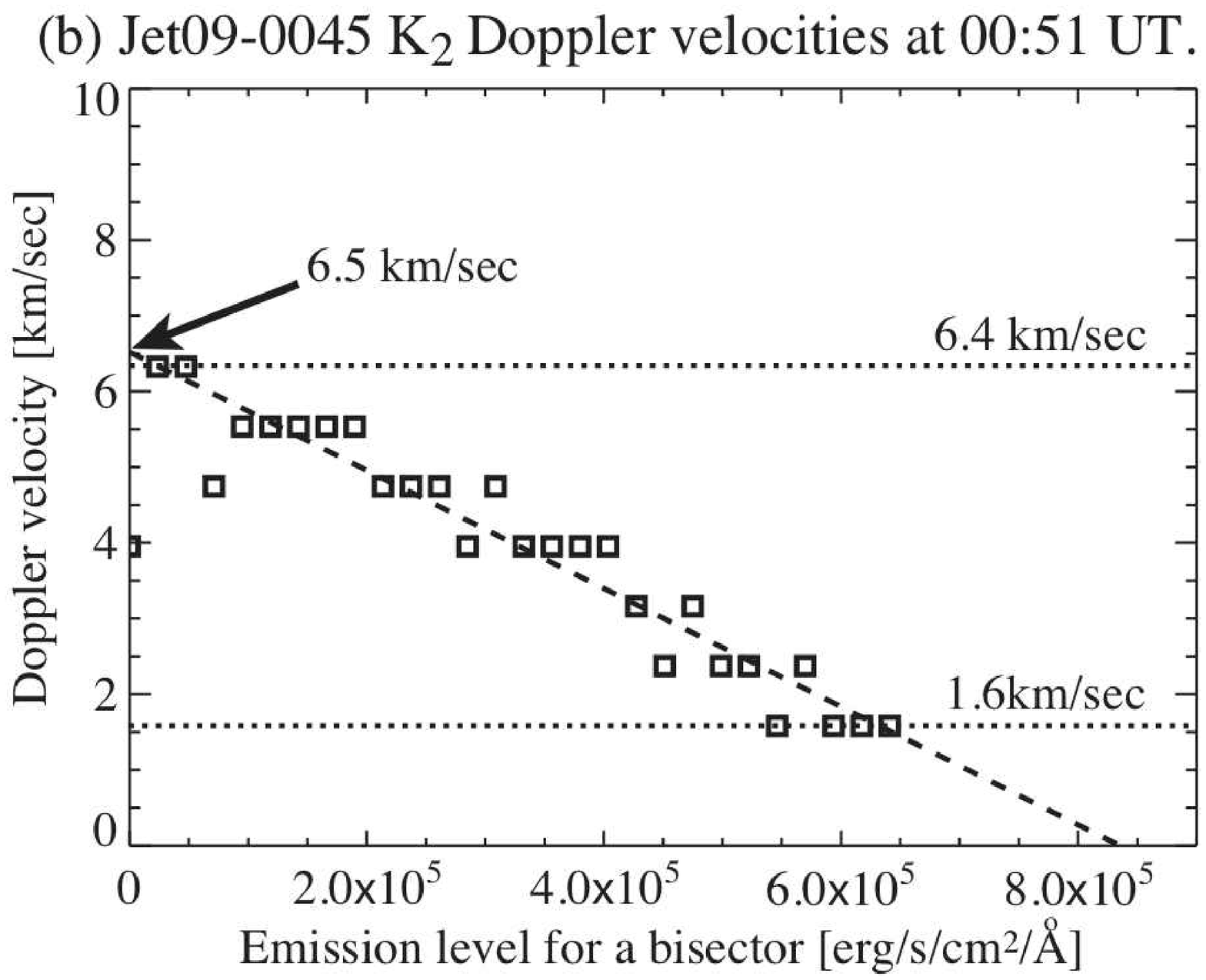}
  \end{center}
  \caption{K$_{2}$ Doppler velocities for (a) ``Jet09-0031'' and
    (b) ``Jet09-0045'': These Doppler velocities are estimated by
    measuring the bisector positions of the two flanks of the emission
    profile in K$_{1}$/K$_{2}$.  Averaged back ground spectra from
    near jet location was subtracted beforehand
    (cf. Fig.~\ref{fig:profiles1}c).  Velocities are measured at
    different intensity levels of an emission curve and plotted in
    these figures.  Lower intensity level for the bisector reflects
    lower atmosphere.}\label{fig:jet1bisec}
\end{figure}

Figure~\ref{fig:evoljet1spec} shows time evolution of the
Ca\emissiontype{II} K spectra of the jets bright cores with the same
timings as mentioned in Figure~\ref{fig:evoljet1}.  The jet bright
cores in this series of jet events show ``red asymmetries'' and were
bright around K$_{1}$.  Since the spectral range of the emission
increases up to + 1.2 \AA\ (in red wing) from the Ca\emissiontype{II}
K line center, we interpreted that this ``red asymmetries'' are not
the apparent shape by the blue shift of the K$_{3}$ absorption line
but the actual red shift of the K$_{2}$ emission itself.  We also
estimated the Doppler velocity of the red ``asymmetries''
(Fig.~\ref{fig:jet1bisec}a, \ref{fig:jet1bisec}b).  
The Doppler velocities (Fig.~\ref{fig:jet1bisec}a and
\ref{fig:jet1bisec}b) are estimated by measuring the bisector
positions of the profiles outside the K$_{2}$ emission peaks.  The
averaged back ground spectra near the jets location were subtracted
before carrying out the velocities measurements
(cf. Fig.~\ref{fig:profiles1}c).  Velocities are measured at different
intensity levels of an emission curve and plotted.  Lower intensity
level corresponds to the lower atmosphere.  The estimated Doppler
velocities of the emission increase in K$_{2}$/K$_{1}$ component are
approximately 5.5 km/s -- 16.4 km/s (red shifted) for an onset of
``Jet09-0031'' and are approximately 1.6 km/s -- 6.4 km/s (red
shifted) for a decay phase of ``Jet09-0045''.
In any case, larger Doppler velocities are found in lower intensity
levels for making the bisector (Fig.~\ref{fig:jet1bisec}a,
\ref{fig:jet1bisec}b).

Figure~\ref{fig:jet1tflux} shows evolutions of the total magnetic flux
and the area of the magnetic sources forming the polarity inversion
line at the footpoints of the jets.  A distinct magnetic flux
cancellation of the negative polarity source is seen at the same time
with the jet occurrence (``Jet09-0045'').  We confirmed that this
distinct flux cancellation actually occurred just below this jet
(Appendix~2.).  The measured magnetic flux cancellation rate and the
decreasing rate of the negative magnetic source area associated with
jets are $2.36\times10^{16}$ Mx/s and $1.17\times10^{14}$ cm$^{2}$/s
for ``Jet09-0045'', respectively.  For ``Jet09-0031'', it is
$1.81\times10^{16}$ Mx/s and $8.25\times10^{13}$ cm$^{2}$/s,
respectively.  Magnetic flux cancellations at the footpoints of the
jets are found to occur intermittently with the jet events in the
chromosphere (Fig.~\ref{fig:jet1tflux}a, \ref{fig:jet1tflux}b).  The
jet activity in this region ceased when all the magnetic flux in
negative polarity source were canceled.

\begin{figure}
  \begin{center}
    \FigureFile(85mm,60mm){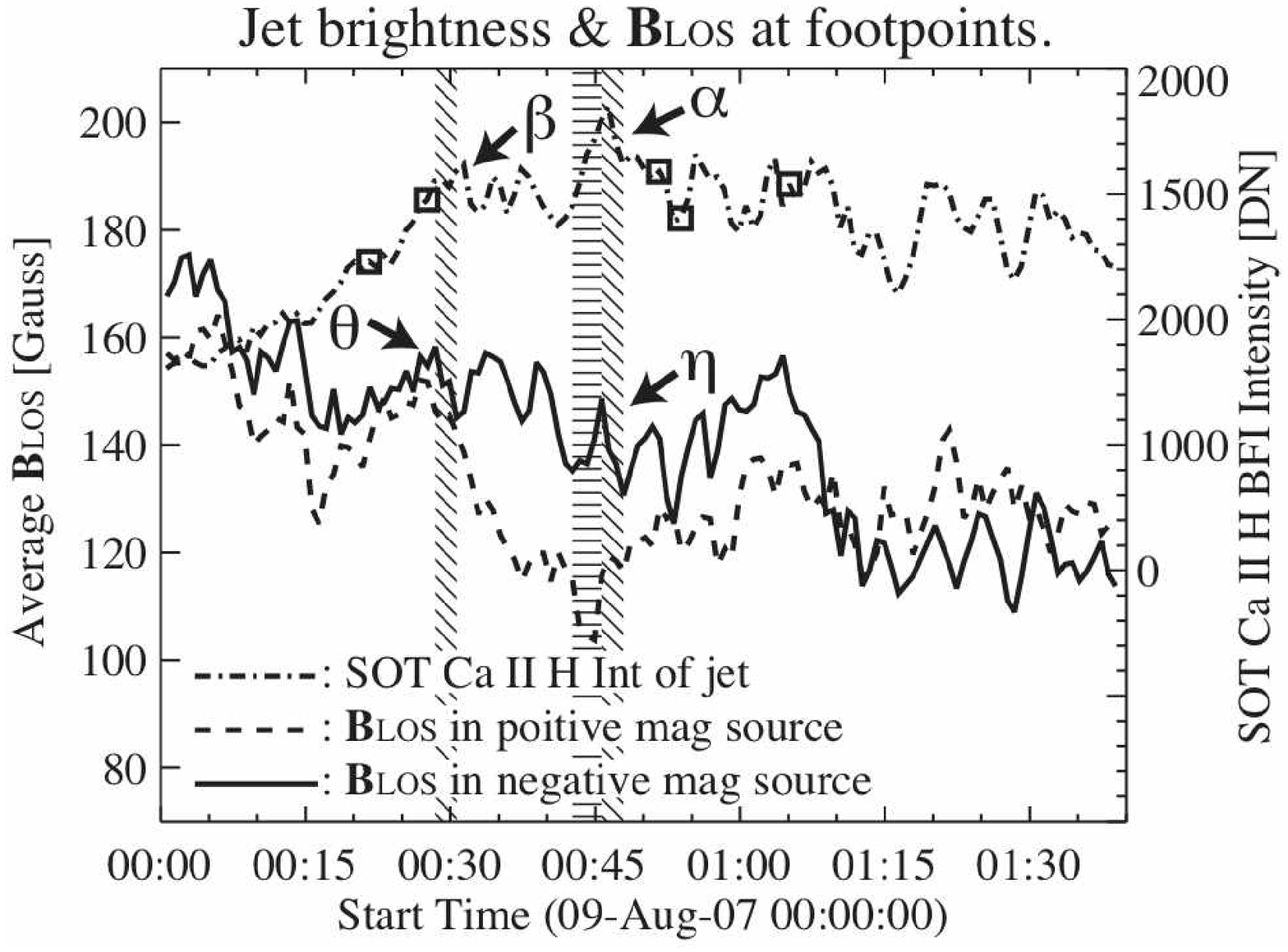}
  \end{center}
  \caption{The same as Fig.~\ref{fig:jet1tflux} but for average
    magnetic intensities of the magnetic sources that involved to the
    series of jet events on August~9 around 00:45 UT: A concentration
    of the negative magnetic flux is seen in the onset phase of
    ``Jet09-0045'' (compare the peaks $\alpha$ and $\eta$ in
    horizontal hatching).  After the jet occurred, the average
    magnetic intensity of the negative magnetic flux decreases with
    the Ca\emissiontype{II}~H intensity decrease (compare $\alpha$ and
    $\eta$ in oblique hatching).  The similar coincidences are fairly
    seen for other peaks (e.g. the peaks $\beta$ and $\theta$).
  }\label{fig:jet1bandc}
\end{figure}

Figure~\ref{fig:jet1bandc} shows an evolution of average magnetic flux
density of the same magnetic sources mentioned in
Fig.~\ref{fig:jet1tflux}.  A concentration of the negative magnetic
flux is seen during the onset of ``Jet09-0045''.  After the jet
occurred, the average magnetic flux density decreased as the
Ca\emissiontype{II}~H intensity decreases.  The similar coincidences
are fairly seen for other Ca\emissiontype{II}~H intensity peaks.

\

\subsubsection{Jets associated with emerging flux (Aug. 8 0333)}

\begin{figure}
  \begin{center}
    \FigureFile(80mm,60mm){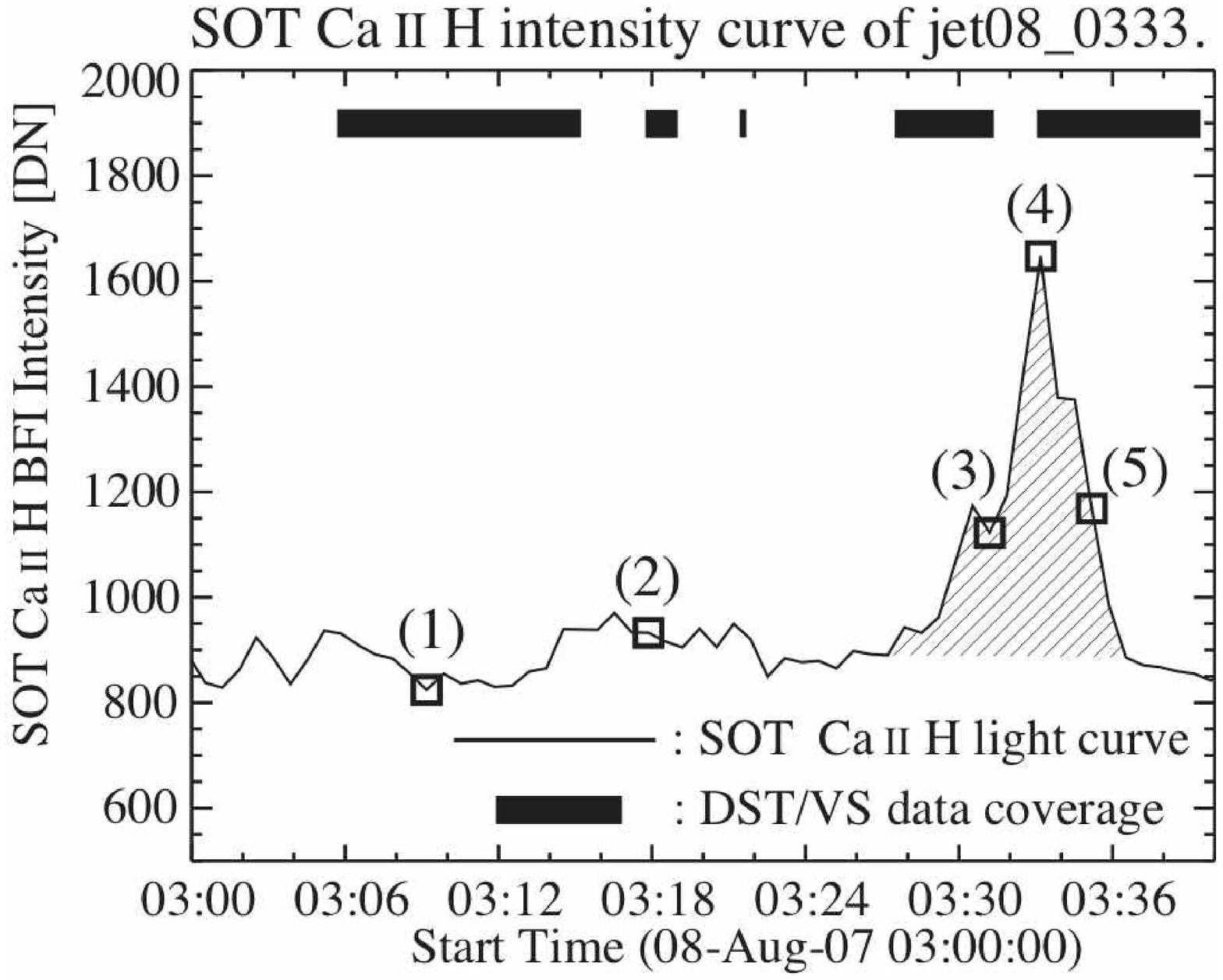}
  \end{center}
  \caption{Time variation of the SOT Ca\emissiontype{II}~H intensity
    (solid curve) at the location where ``Jet08-0333'' bright core
    appeared: The strong peak (oblique hatching) corresponds to
    ``Jet08-0333'' (see Table~1).  Sampled area is 1.\arcsec09 $\phi$
    ($\sim$800km $\phi$) disk, and the intensities are averaged in it
    (in arbitrary unit).  Tracking for the differential rotation of
    the sun is applied.  Each square with a number on this curve
    corresponds to the timing of a row in Fig.~\ref{fig:evoljet2} and
    of a Ca\emissiontype{II} K spectrum in
    Fig.~\ref{fig:evoljet2spec}, with the corresponding number,
    respectively.  The horizontal black bars near the top of this
    figure show the DST/VS data coverage.}\label{fig:jet2lcurve}
\end{figure}

In this section we discuss a single jet event that occurred just near
an emerging flux region.  Figure~\ref{fig:jet2lcurve} shows a SOT
Ca\emissiontype{II} H light curve for ``Jet08-0333'' (see Table~1).
The strong peak at the end of this light curve corresponds to this
jet.  The size of the jet bright core and the maximum length of this
jet in SOT Ca\emissiontype{II} H images are 2--3 times larger and
longer than the jets in section~3.2.1. (see Table~1).

\begin{figure*}
  \begin{center}
    \FigureFile(150mm,165mm){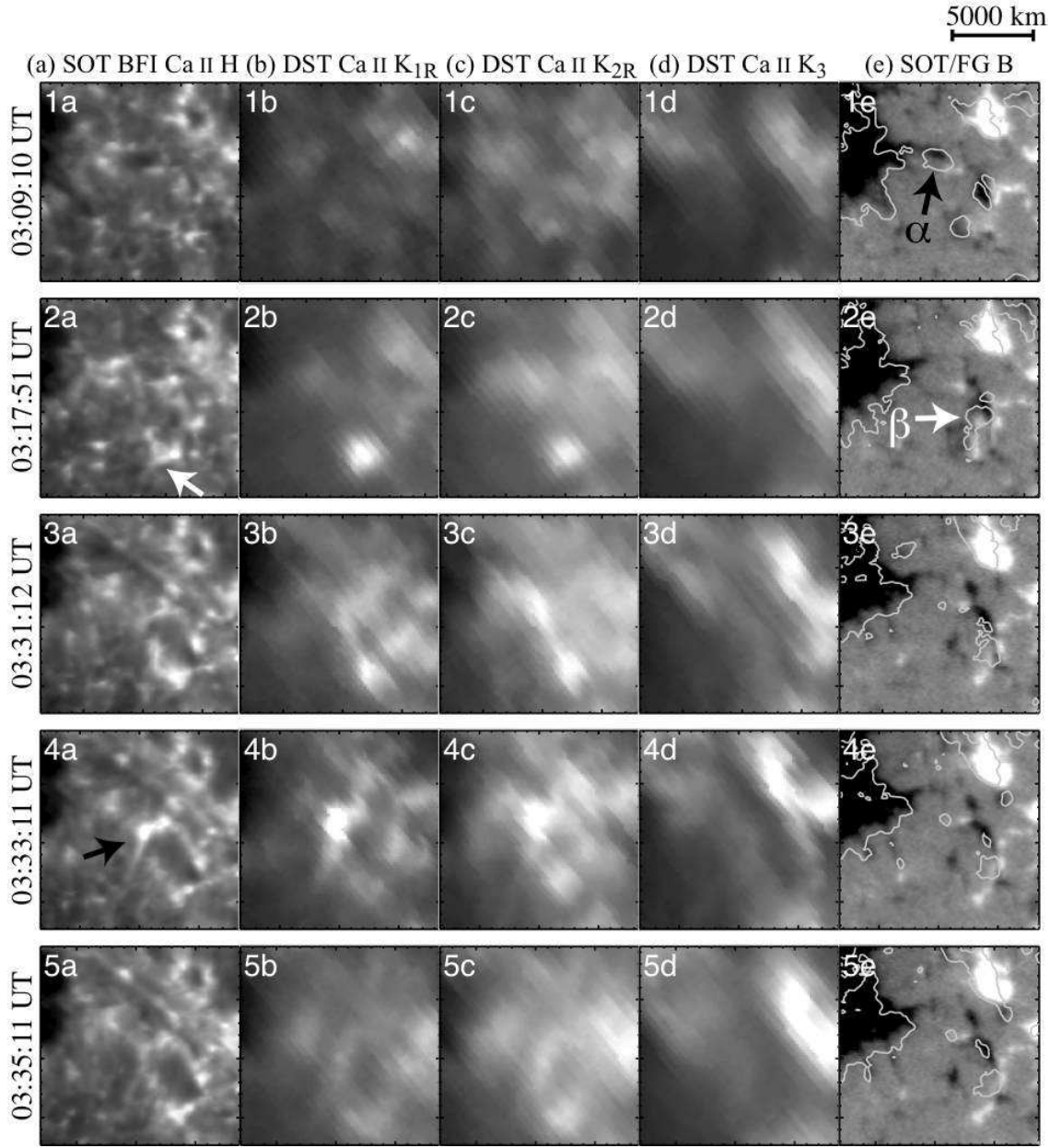}
  \end{center}
  \caption{Evolution of ``Jet08-0333'' with DST/Hida
    Ca\emissiontype{II}~K spectroheliograms and SOT/Hinode images:
    Details are the same as in Fig.~\ref{fig:evoljet1} but for
    ``Jet08-0333''.  Each image has the same FOV of
    16.\arcsec8$\times$16.\arcsec8.  Each row in this figure
    corresponds to a square in Fig.~\ref{fig:jet2lcurve} and a
    Ca\emissiontype{II}~K spectra in Fig.~\ref{fig:evoljet2spec}, with
    the corresponding number.  The 3rd, 4th, and 5th rows represent
    the onset, peak, and decay phase of ``Jet08-0333'', respectively.
    A jet of the bottom direction having a bright cusp at its
    footpoint is clearly seen in frame (4a) (shown black arrow).  This
    jet appeared above the positive magnetic source of an emerging
    dipole magnetic source ``$\alpha$'' in frame (1e) but initiated by
    the extension of another larger emerging dipole ``$\beta$'' in
    frame (2e).  The bright cusp in frame (2a) (shown white arrow) is
    the footpoint of another chromospheric jet ``Jet08-0319'' (see
    Table~1).  This jet occurred at the other end of the emerging
    dipole ``$\beta$'', and traveled to the bottom direction, same as
    ``Jet08-0333''.}\label{fig:evoljet2}
\end{figure*}

Figure~\ref{fig:evoljet2} shows the evolution of ``Jet08-0333'' with
the SOT BFI and NFI and the DST/Hida spectroheliograms.  The onset,
peak, and decay phase of ``Jet08-0333'' is shown.  A jet having a
bright cusp toward the south is clearly seen in frame (4a) (shown
black arrow).  This jet appeared near the positive magnetic source of
an emerging dipole ``$\alpha$'' in frame (1e) but initiated by the
extension of another larger emerging dipole ``$\beta$'' in frame (2e).
Another bright cusp in frame (2a) (shown white arrow) shows the jet
``Jet08-0319'' (see Table~1).  This jet occurred at the other end of
the emerging dipole ``$\beta$'', and was ejected into a southward
direction, same as ``Jet08-0333''.  In any case, the jets or jet
bright cores are bright at K$_{1}$ and K$_{2}$.  The K$_{1}$ images
show better contrast jet structures.

\begin{figure}
  \begin{center}
    \FigureFile(80mm,145mm){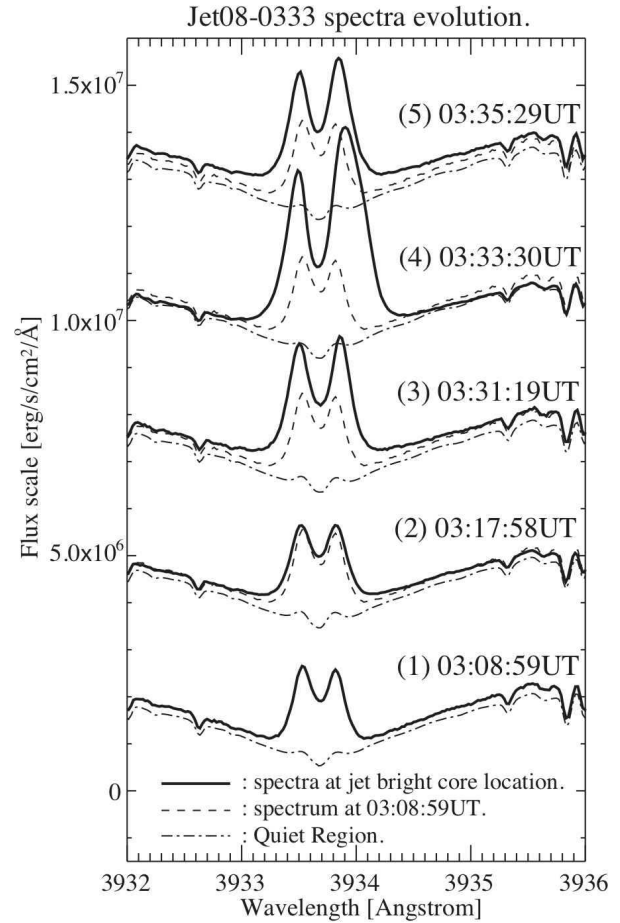}
  \end{center}
  \caption{Evolution of the Ca\emissiontype{II}~K spectra of the jet
    bright core of ``Jet08-0333'': Details are the same as
    Fig.~\ref{fig:evoljet1spec} but for ``Jet08-0333''.  The solid
    curves indicate the spectra of the jet bright core ((3): jet
    onset, (4): peak, and (5): decay phase), or those of the same
    location but rather in quiet condition ((1) and (2)).  A reference
    spectrum (broken curves; the spectrum from 03:08:59 UT at the same
    location) is over plotted for showing intensity increase with the
    jet.  This ``Jet08-0333'' is bright at K$_{1}$, same as the jets
    ``Jet09-0031'' and ``Jet09-0045'' in Fig.~\ref{fig:evoljet1spec}.
    The ``red asymmetry'' is obvious only at the peak
    phase.}\label{fig:evoljet2spec}
\end{figure}

\begin{figure}
  \begin{center}
    \FigureFile(80mm,56mm){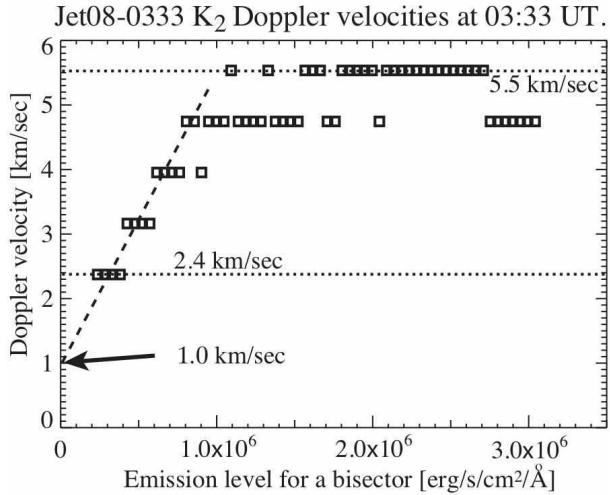}
  \end{center}
  \caption{K$_{2}$ Doppler velocities for ``Jet08-0333'': Details are
    the same as in Fig.~\ref{fig:jet1bisec}.}\label{fig:jet2bisec}
\end{figure}

Figure~\ref{fig:evoljet2spec} shows the time evolution of
Ca\emissiontype{II}~K spectra of ``Jet08-0333''.  ``Jet08-0333'' is
bright at K$_{1}$, same as the jets ``Jet09-0031'' and ``Jet09-0045''
in Fig.~\ref{fig:evoljet1spec}.  The ``red asymmetry'' is obvious at
the peak phase.  Estimated K$_{2}$ Doppler velocities for the ``red
asymmetry'' using the bisector positions of the emission profile (cf.
section~3.2.1) are approximately 2.4 km/s -- 5.5 km/s red shifted.
Larger Doppler velocities are found in higher emission levels for the
bisector, in contrast to the jets mentioned in section~3.2.1
(Fig.~\ref{fig:jet2bisec}).

\begin{figure}
  \begin{center}
    \FigureFile(85mm,60mm){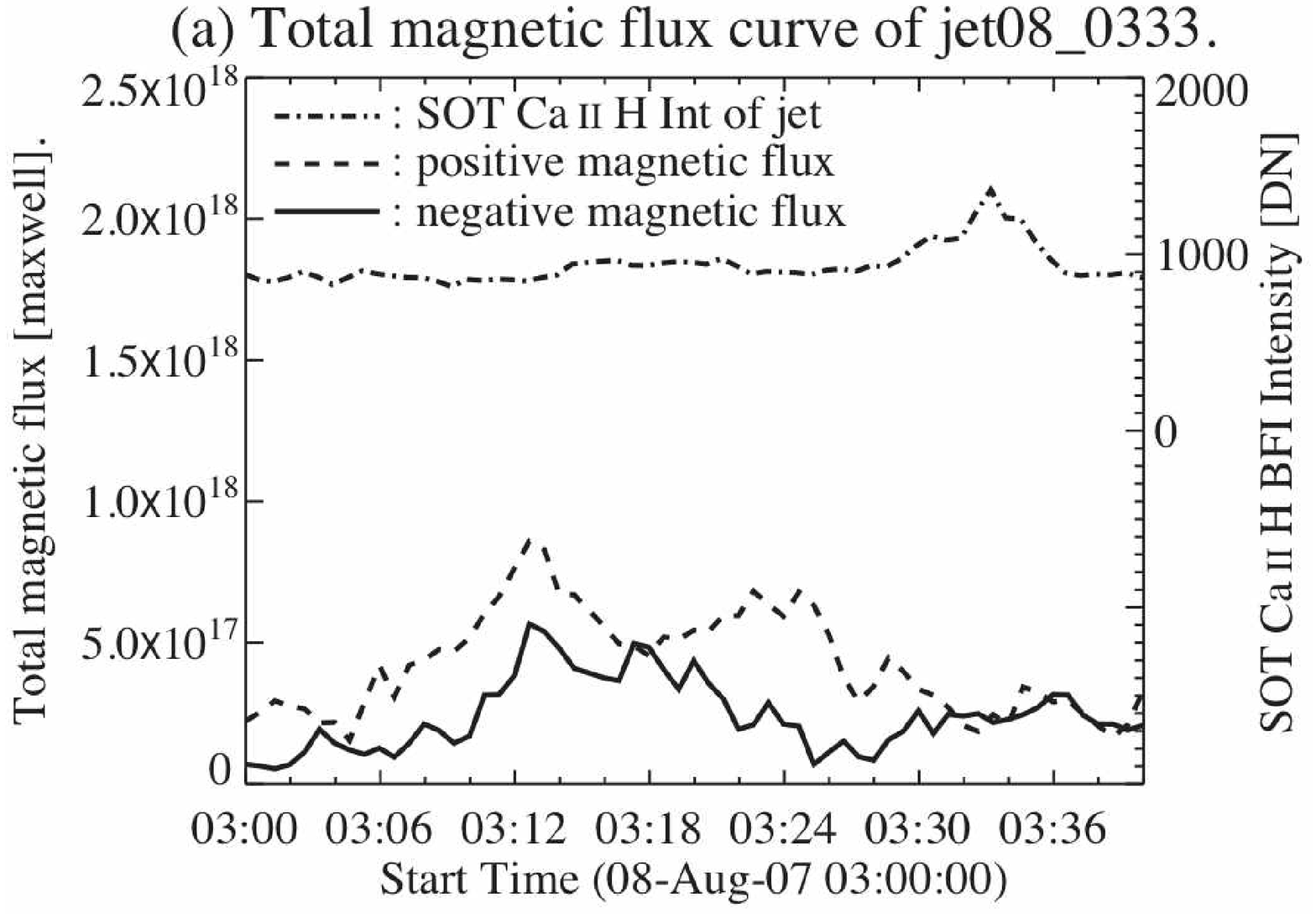}
    \FigureFile(85mm,60mm){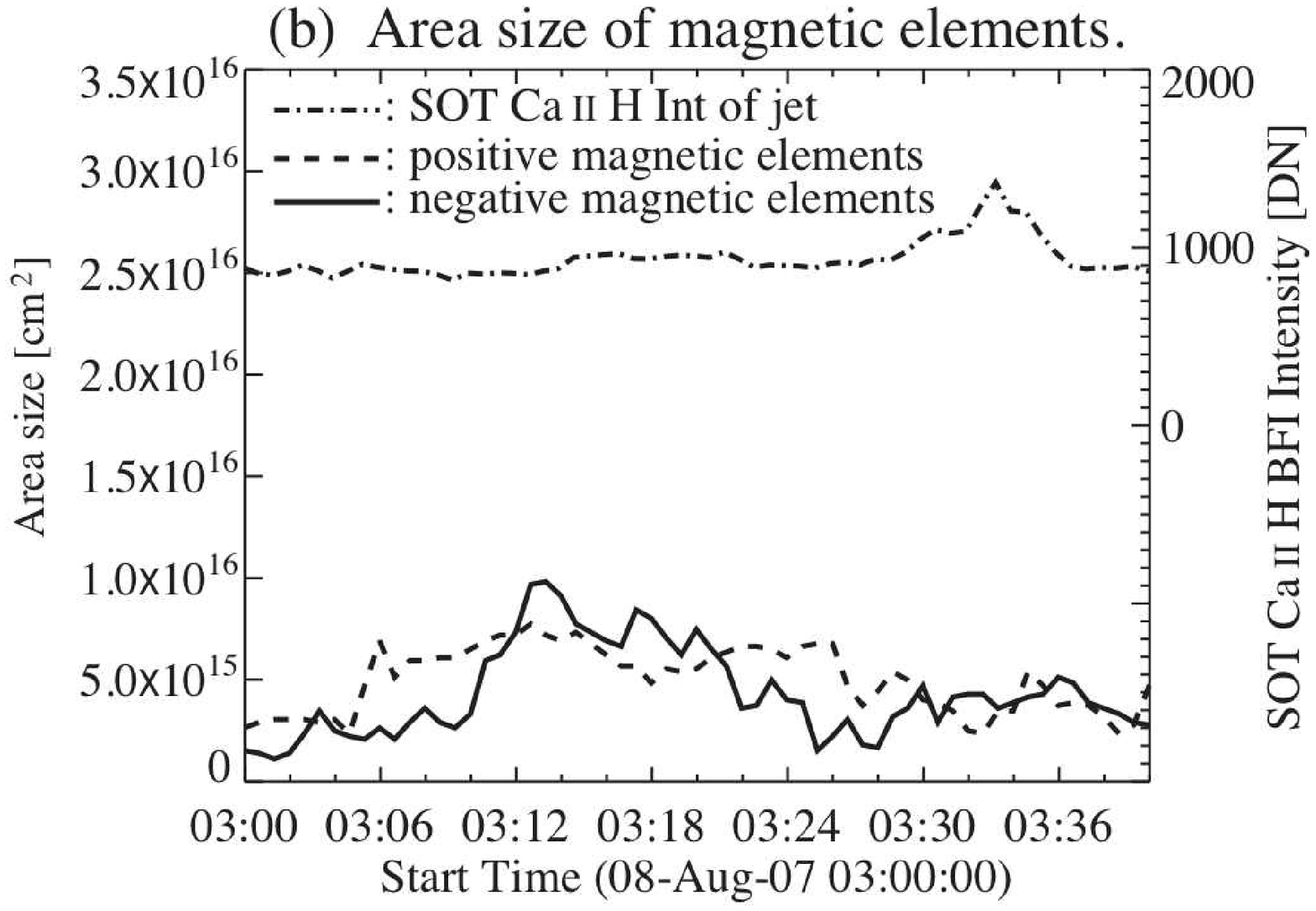}
  \end{center}
  \caption{Evolutions of (a) total magnetic flux and (b) area size of
    the magnetic sources that involved to ``Jet08-0333'': Details are
    the same as in Fig.~\ref{fig:jet1tflux} but for ``Jet08-0333''.
    Samplings are made for the area of 3.\arcsec84 $\phi$
    ($\sim$2800km $\phi$) disk, centered at the location of the jet
    bright core.  Tracking for the differential rotation of the sun is
    applied.  No distinct magnetic flux cancellation is seen at the
    same timing of the ``Jet08-0333'', but a cancellation of the
    positive magnetic flux is seen about 8 minutes before the jet
    peak.}\label{fig:jet2tflux}
\end{figure}

Figure~\ref{fig:jet2tflux} shows evolution of total magnetic flux and
area of the magnetic sources involved in the ``Jet08-0333''.  No
distinct magnetic flux cancellation is seen during the same time of
the ``Jet08-0333'', but a cancellation of the positive magnetic flux
of an emerging dipole ``$\alpha$'' (see Fig.~\ref{fig:evoljet2}(1e))
is seen about 8 minutes before the Ca\emissiontype{II}~H jet.

\

\subsubsection{Jets associate with MMFs (Aug. 8 0046)}

\begin{figure}
  \begin{center}
    \FigureFile(80mm,60mm){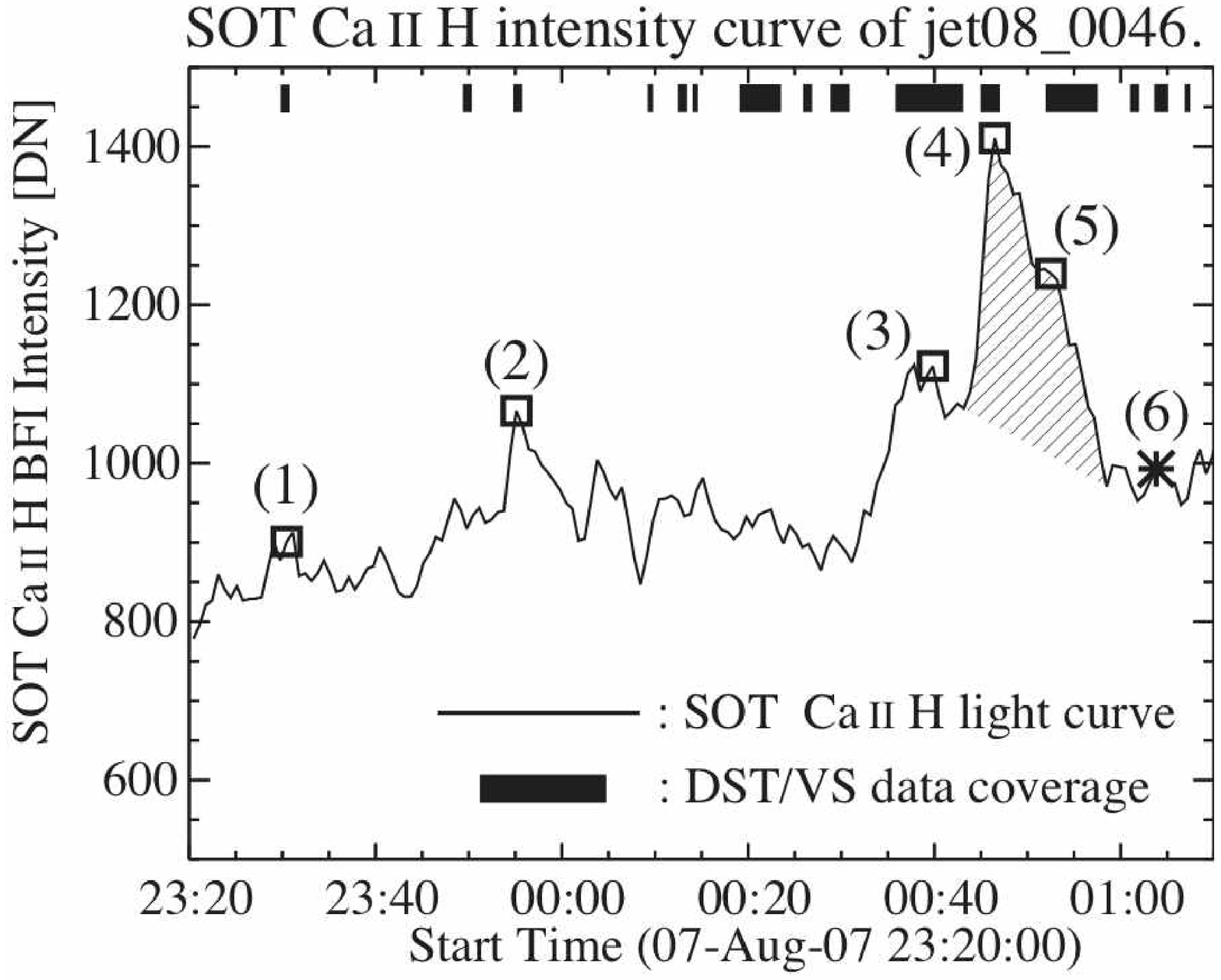}
  \end{center}
  \caption{Time variation of the SOT Ca\emissiontype{II}~H intensity
    at the location on a moving magnetic feature (MMF) where
    ``Jet08-0046'' bright core appeared: The strong peak (oblique
    hatching) corresponds to the ``Jet08-0046'' (see Table~1).  This
    intensity curve was made as tracking the locations of a MMF, which
    was traveling (shown white arrows in column (e) in
    Fig.~\ref{fig:evoljet3}). Sampled area is 2.\arcsec18 $\phi$
    ($\sim$1600km $\phi$) disk, and the intensities are averaged in it
    (in arbitrary unit).  The squares with the numbers ((1)--(5)) on
    this curve correspond to the timings of rows in
    Fig.~\ref{fig:evoljet3} and of Ca\emissiontype{II}~K spectra in
    Fig.~\ref{fig:evoljet3spec}, with the corresponding numbers,
    respectively.  The asterisk, indicated by (6), represents the time
    of the reference Ca\emissiontype{II}~K spectra in
    Fig.~\ref{fig:evoljet3spec} (broken curves).  The horizontal black
    bars near the top of this figure show the DST/VS data coverage.
  }\label{fig:jet3lcurve}
\end{figure}

In the previous sections~3.2.1~and~3.2.2, chromospheric anemone jets
occurred on a polarity inversion line that is involved in one or two
emerging fluxes (EFR).  In this section we introduce a different case,
i.e. a chromospheric anemone jet on a polarity inversion line formed
between a moving magnetic feature (MMF) and satellite magnetic patches
of opposite polarity.

Figure~\ref{fig:jet3lcurve} shows time variation of the SOT
Ca\emissiontype{II}~H intensities by tracking the locations of an MMF
where ``Jet08-0046'' bright core appeared.  The strong peak around
00:40 UT corresponds to the ``Jet08-0046'' (see Table~1).  The life
time of the ``bright core'' of this jet is around 320 seconds, but the
jet itself seems to have shorter life time (cf. Table~1 for more
details).

\begin{figure*}
  \begin{center}
    \FigureFile(145mm,179mm){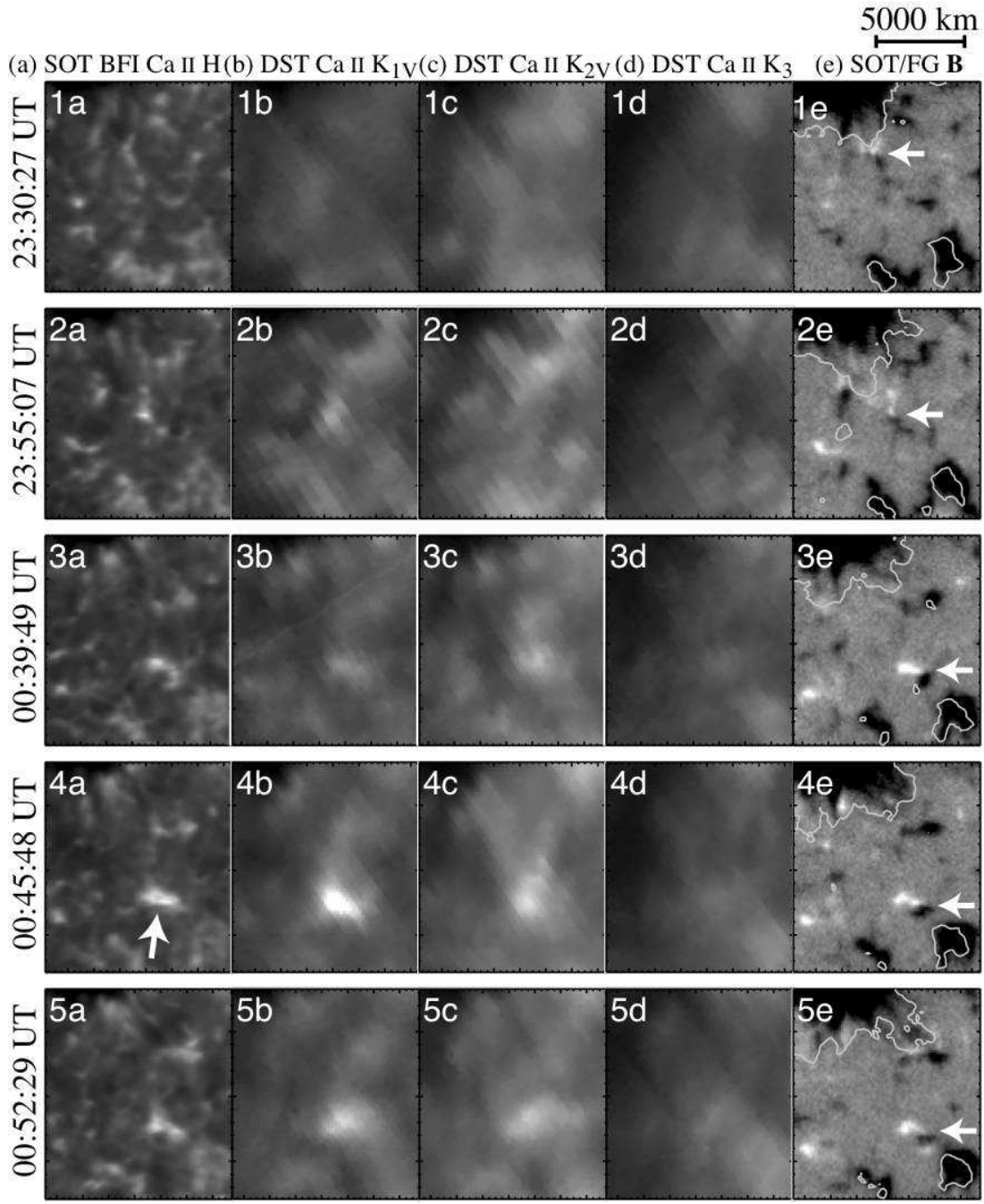}
  \end{center}
  \caption{Evolution of ``Jet08-0046'' with DST/Hida
    Ca\emissiontype{II}~K spectroheliograms and SOT/Hinode images:
    Details are the same as in Fig.~\ref{fig:evoljet1} but for
    ``Jet08-0046'', and Ca\emissiontype{II}~K$_{1V}$ and
    Ca\emissiontype{II}~K$_{2V}$ are chosen instead of the red wing
    spectroheliograms.  Each image has the same FOV of
    14.\arcsec4$\times$16.\arcsec2.  Each row in this figure
    corresponds to a square in Fig.~\ref{fig:jet3lcurve} and a
    Ca\emissiontype{II}~K spectra in Fig.~\ref{fig:evoljet3spec}, with
    the corresponding number.  The 3rd, 4th, and 5th rows represent
    the onset, peak, and decay phase of ``Jet08-0046'', respectively.
    A jet of the leftward direction is seen in frame (4a) (shown white
    arrow).  This jet appeared above a southward traveling MMF, which
    had left the south edge of the sunspot (shown white arrows in
    frames (1e)--(5e)).  This MMF seemingly collided with other
    magnetic elements, then the jet occurred.}\label{fig:evoljet3}
\end{figure*}

Figure~\ref{fig:evoljet3} shows the evolution of ``Jet08-0046'' with
the SOT BFI and NFI and the DST/Hida spectroheliograms.  In this
figure, blue wing images, K$_{1V}$ and K$_{2V}$, are chosen instead of
the red wing that is chosen in previous figures since it is brighter
in the blue wing of the Ca\emissiontype{II} K line.  A jet of the
leftward direction is fairly seen in frame (4a) (see white arrow for
the jet bright core).  This jet occurred above a southward traveling
MMF, which had left the south edge of the sunspot.  This MMF seemingly
collided with other magnetic elements, and then the jet occurred.

\begin{figure}
  \begin{center}
    \FigureFile(80mm,143mm){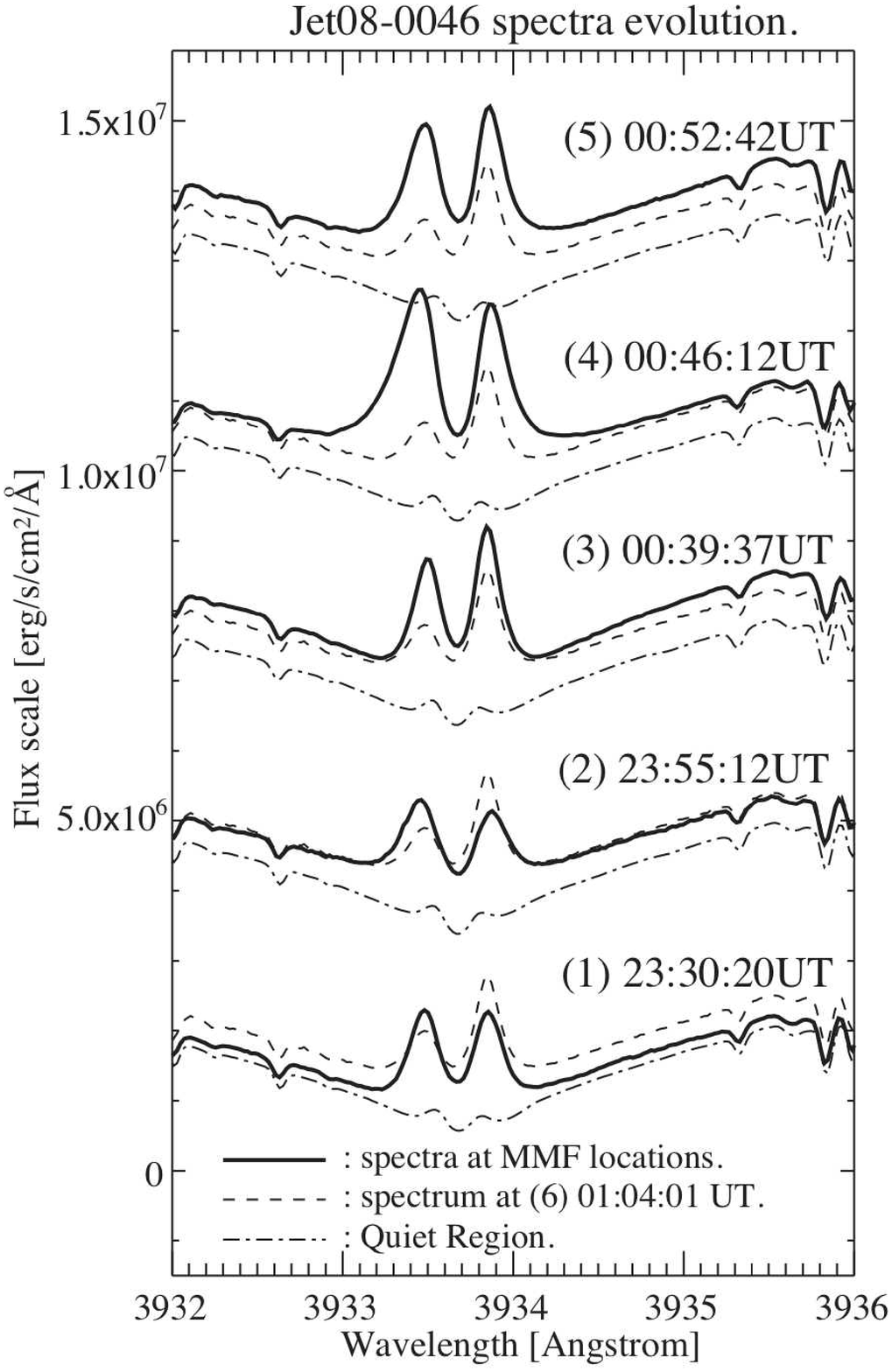}
  \end{center}
  \caption{Evolution of Ca\emissiontype{II}~K spectra of the jet
    bright core of ``Jet08-0046'': Details are the same as in
    Fig.~\ref{fig:evoljet1spec} but for ``Jet08-0046''.  The solid
    curves indicate the spectra of the jet bright core ((3): jet
    onset, (4): peak, and (5): decay phase), or those of the locations
    on the MMF with tracking but rather in quiet condition ((1) and
    (2)).  A reference spectrum (broken curves; the spectrum from
    01:04:01 UT at the MMF's location) is over plotted for showing
    intensity increases with jets.  This reference spectrum is chosen
    since the MMF was traveling before the jet occurred.  In contrast
    to the spectra of the jets with emerging dipole magnetic sources
    (``Jet08-0333'', ``Jet09-0031'', and ``Jet09-0045''),
    ``Jet08-0046'' is bright mainly at K$_{1V}$, and ``blue
    asymmetry''.}\label{fig:evoljet3spec}
\end{figure}

\begin{figure}
  \begin{center}
    \FigureFile(80mm,56mm){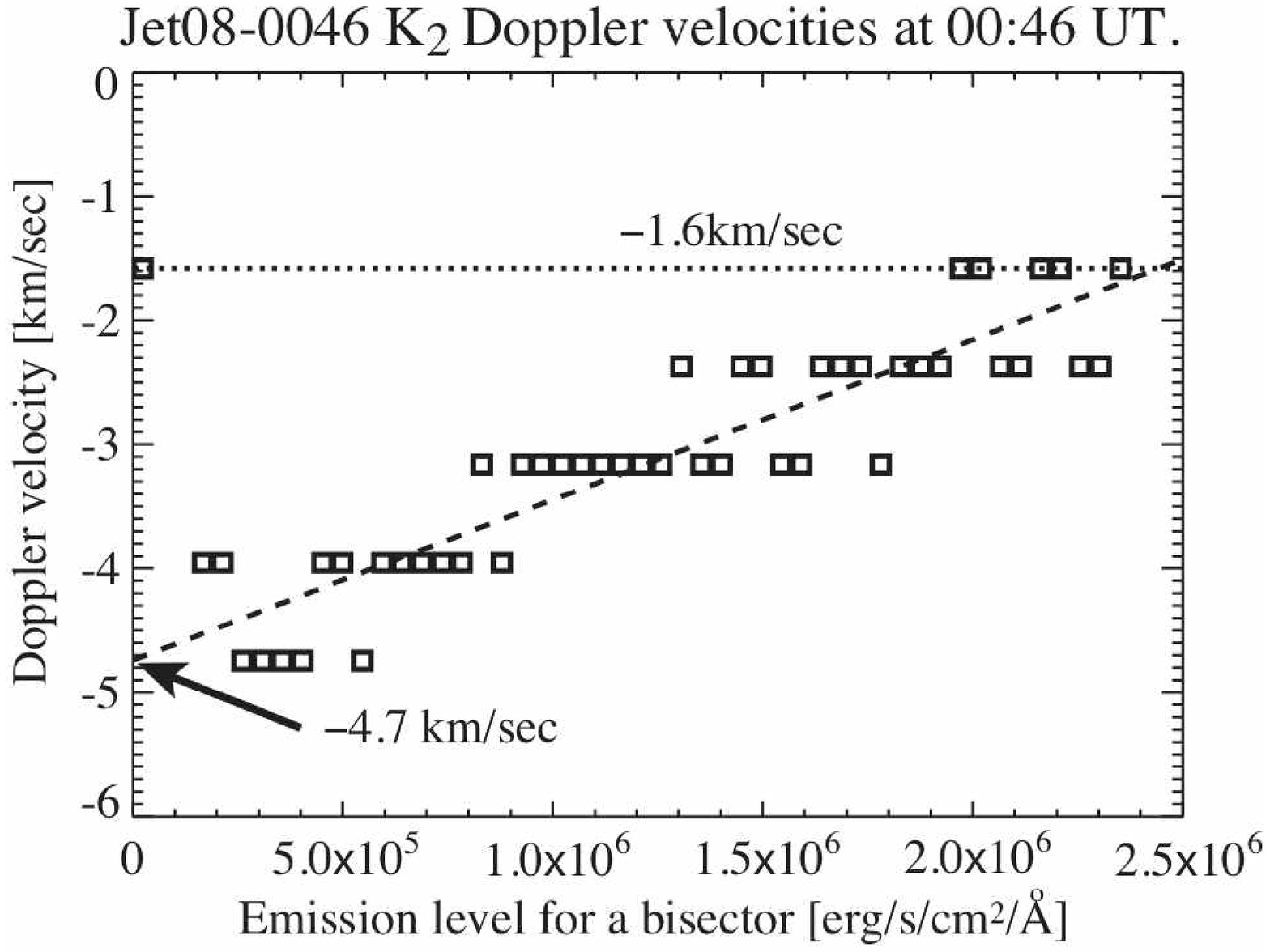}
  \end{center}
  \caption{K$_{2}$ Doppler velocities for ``Jet08-0046'': Details are
    the same as in Fig.~\ref{fig:jet1bisec}.  This event shows blue
    shift in K$_{2}$ component.}\label{fig:jet3bisec}
\end{figure}

Figure~\ref{fig:evoljet3spec} shows time evolution of the
Ca\emissiontype{II} K spectra of the jet.  In contrast to the spectra
of the jets with emerging dipole magnetic sources (``Jet08-0333'',
``Jet09-0031'', and ``Jet09-0045''), ``Jet08-0046'' shows ``blue
asymmetry''.  Estimated K$_{2}$ blue shifted Doppler velocities for
the ``blue asymmetry'' are approximately 1.6 km/s -- 4.7 km/s.  Larger
Doppler velocities are found in the lower emission levels for the
bisector, same as the jets mentioned in section~3.2.1 (see
Fig.~\ref{fig:jet3bisec}).

\begin{figure}
  \begin{center}
    \FigureFile(85mm,60mm){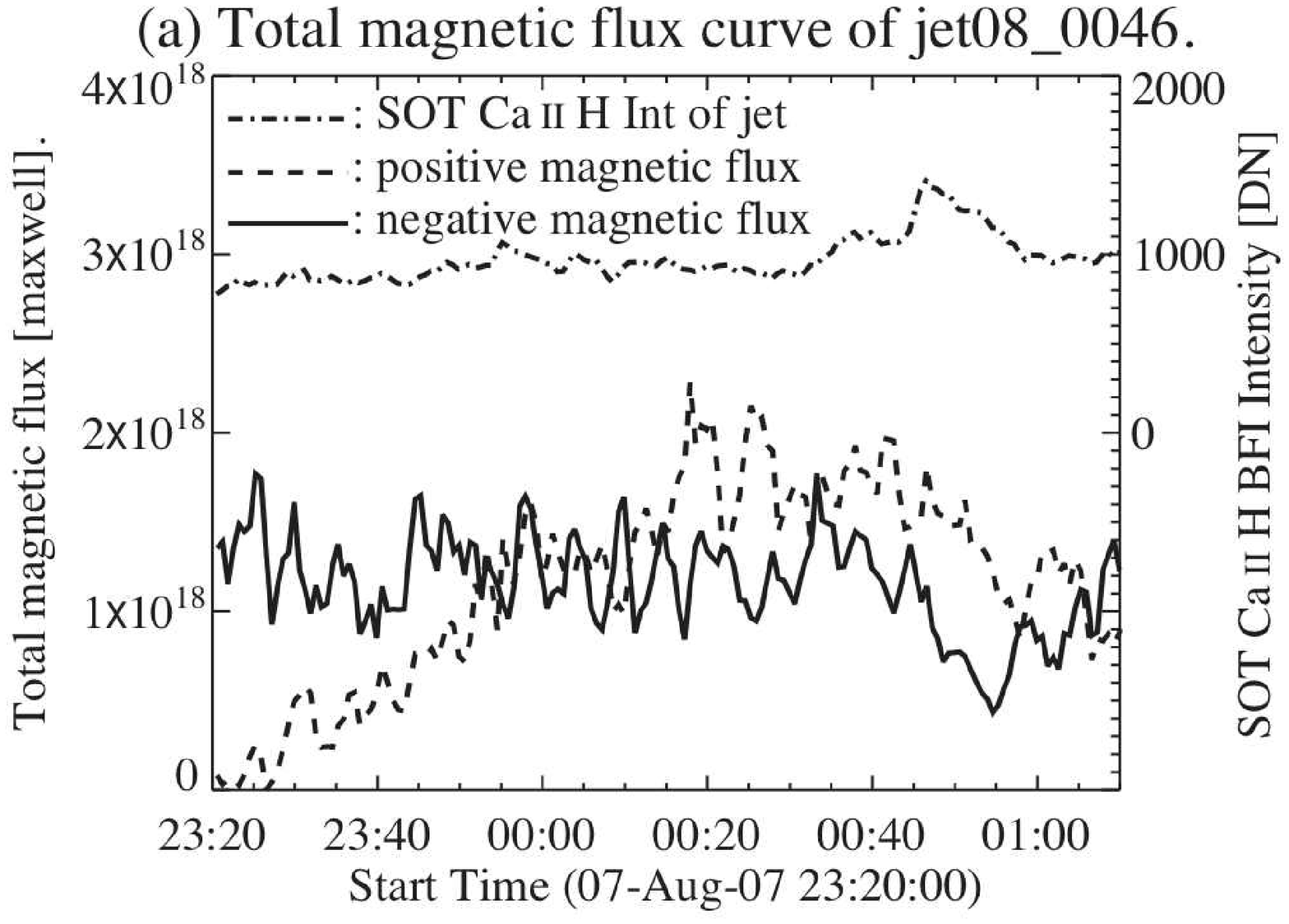}
    \FigureFile(85mm,60mm){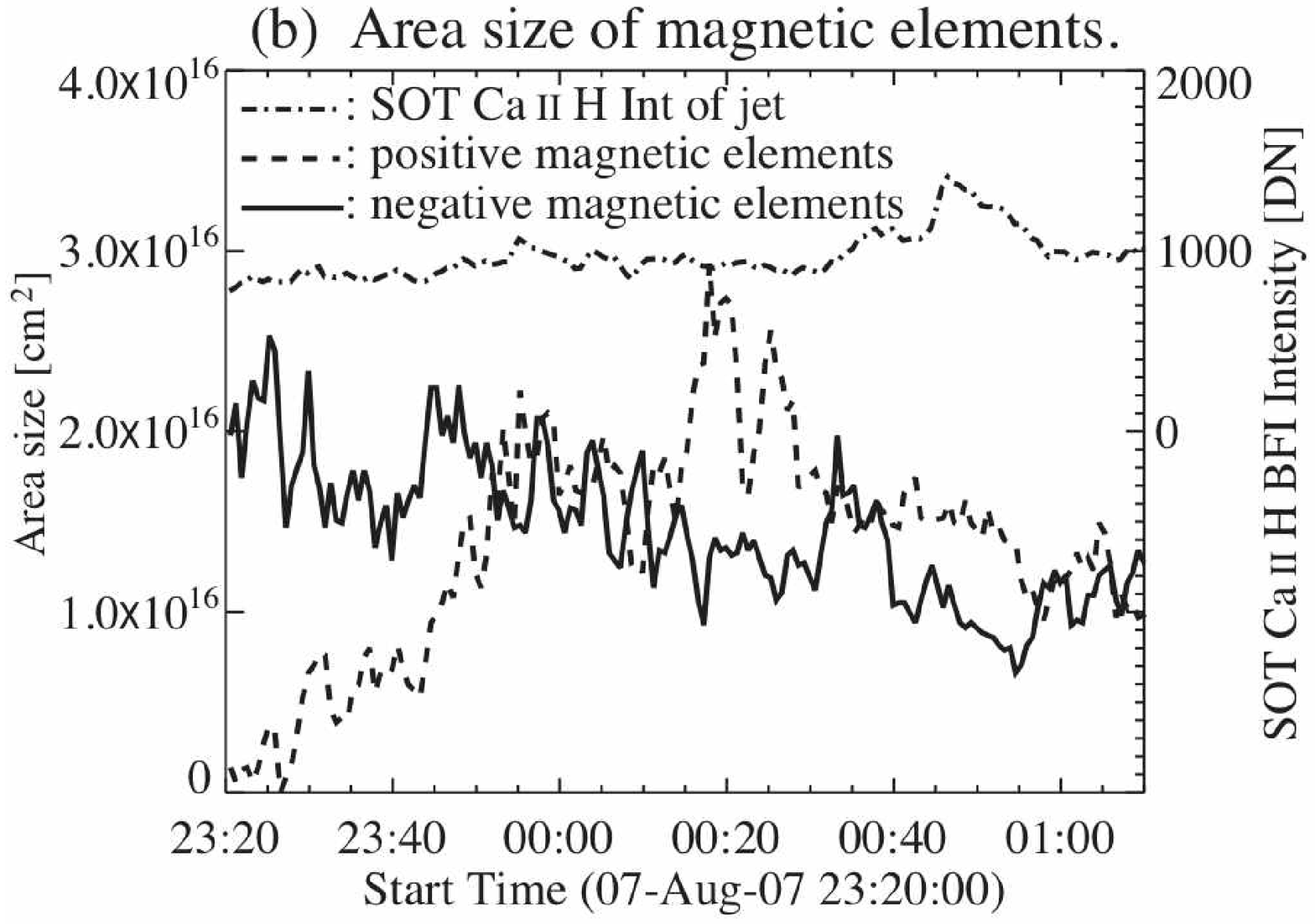}
  \end{center}
  \caption{Evolutions of (a) total magnetic flux and (b) area size of
    the magnetic sources around the locations of the MMF that involved
    to ``Jet08-0046'': Details are the same as in
    Fig.~\ref{fig:jet1tflux} but for ``Jet08-0046''.  Samplings are
    made for the area of 6.\arcsec4 $\phi$ ($\sim$4700km $\phi$) disk,
    centered at the locations of the MMF with tracking it.  A
    cancellation of the negative magnetic flux is seen around the same
    timing of the ``Jet08-0046''.  This negative magnetic flux
    involved in the satellite magnetic patches of negative polarity
    collided with the MMF (cf.
    Fig.~\ref{fig:evoljet3}).}\label{fig:jet3tflux}
\end{figure}

Figure~\ref{fig:jet3tflux} shows evolutions of total magnetic flux and
area of the magnetic sources that is involved in the ``Jet08-0046''.
Samplings are made for the area of 6.\arcsec4 $\phi$ ($\sim$4700km
$\phi$) disk, centered at the locations of the MMF with tracking it.
A cancellation of the negative magnetic flux is seen simultaneously
with the ``Jet08-0046'' event.  This negative magnetic flux involved
in the satellite magnetic patches of negative polarity collided with
the MMF (see Fig.~\ref{fig:evoljet3}).

\section{Discussions}

\subsection{Estimates of energy release rate}

In this section we estimate the energy release rate of the
chromospheric jets by two different methods, (1) from the magnetic
cancellation rate assuming magnetic reconnection, and (2) from the
emission increase around the K$_{2}$/K$_{1}$ on the
Ca\emissiontype{II} K line profiles.
By assuming two dimensional steady magnetic reconnection, the released
magnetic energy can be estimated as the Poynting flux entering from
both sides into the reconnecting region using the relation,
\begin{equation}
\frac{dE_{\mathrm{mag}}}{dt} =
2\,\frac{B_{\mathrm{ch}}^2}{4\pi}\,v_{\mathrm{in}}\,H\,l_{\mathrm{PIL}},
\end{equation}
where dE$_{\mathrm{mag}}/dt$ is the magnetic energy release rate due
to magnetic reconnection in the lower chromosphere, $B_{\mathrm{ch}}$
is magnetic flux density in the lower chromosphere, $v_{\mathrm{in}}$
is an inflow velocity to the reconnection site, $H$ ($\approx 150$ km)
is the pressure scale height in the lower chromosphere, and
$l_{\mathrm{PIL}}$ is the length of magnetic polarity inversion line
where the jet has occurred.  Here we assumed that the vertical size of
the reconnection region is approximately the same size as the diameter
of flux tubes and that is the pressure scale height $H$.
Since we cannot know the actual inflow velocity $v_{\mathrm{in}}$ in
the lower chromosphere, we use a horizontal converging velocity at the
photosphere as an assumption.  That is given by
\begin{equation}
v_{\mathrm{cancel}} = \frac{dS}{dt}{l_{\mathrm{PIL}}}^{-1},
\end{equation}
where $v_{\mathrm{cancel}}$ is the horizontal converging velocity to
the magnetic polarity inversion line at the photosphere, $dS/dt$ is
decreasing rate of area of the magnetic source at the footpoint of a
jet.
Since we have only the photospheric magnetic field observation, we
estimate the chromospheric value from the following relation,
\begin{equation}
\frac{B_{\mathrm{ch}}}{B_{\mathrm{ph}}} = e^{-\frac{Z}{2H}},
\end{equation}
where $B_{\mathrm{ph}}$ is a magnetic flux density in the photosphere,
and $Z$ is height of the reconnection site from the photosphere.
We use the following relation for $B_{\mathrm{ph}}$ as,
\begin{equation}
B_{\mathrm{ph}} = \frac{d\Phi_{\mathrm{ph}}}{dt}\,(\frac{dS}{dt})^{-1},
\end{equation}
where $d\Phi_{\mathrm{ph}}/{dt}$ is an observed magnetic flux cancellation rate.
Here $B_{\mathrm{ph}}$ is the spatially averaged value of the observed
magnetic cancellation rate.
We consider the ``filling factor'' $f$,
\begin{equation}
\overline{B_{\mathrm{ph}}} = f\,B_{\mathrm{ph}},
\end{equation}
\begin{equation}
S = f\,\overline{S},
\end{equation}
where $\overline{B_{\mathrm{ph}}}$ and $\overline{S}$ are observed,
and $B_{\mathrm{ph}}$ and $S$ are intrinsic magnetic flux density and
area, respectively.  The observed magnetic flux is $\Phi_{\mathrm{ph}}
= B_{\mathrm{ph}}\,S = \overline{B_{\mathrm{ph}}}\,\overline{S}$.  It
is not affected by the filling factor.
Finally, we get the equation for magnetic energy release rate due to
magnetic reconnection in the lower chromosphere as,
\begin{equation}
\frac{dE_{\mathrm{mag}}}{dt} = 2\,\frac{H\,e^{-\frac{Z}{H}}}{4\pi\,f}\,
(\frac{d\Phi_{\mathrm{ph}}}{dt})^2\,
(\frac{d\overline{S}}{dt})^{-1}.
\end{equation}
Now we apply the formula to estimate the energy release rate of two
jets, i.e.  ``Jet09-0045'' and ``Jet09-0031''.  We use $f = 0.15$,
assuming that the actual magnetic flux density of each magnetic
element $B_{\mathrm{ph}} \approx 1000$ Gauss, for the observed average
magnetic flux density $\overline{B_{\mathrm{ph}}}$ = 150 Gauss for
``Jet09-0045'' ($\eta$ in Fig.~\ref{fig:jet1bandc}).
For ``Jet09-0045'', we have $d\Phi_{\mathrm{ph}}/dt =
-2.36\times10^{16}$ Mx/s ($\gamma$ in Fig.~\ref{fig:jet1tflux}),
$d\overline{S}/dt = -1.17\times10^{14}$ cm$^{2}$/s ($\epsilon$ in
Fig.~\ref{fig:jet1tflux}), $l_{\mathrm{PIL}} = 9.50\times10^{7}$ cm,
and in the lower chromosphere or upper photosphere H $\approx 150$ km.
Thus assuming the height of the reconnection site for the jet as $Z
\approx 600$ km (lower chromosphere) or $Z \approx 300$ km (upper
photosphere), we obtain from Eq.~(7), $dE_{\mathrm{mag}}/dt =
(1.4-10)\times10^{24}$\,erg/s.  For ``Jet09-0031'', we obtain
$dE_{\mathrm{mag}}/dt = (1.1-8.1)\times10^{24}$\,erg/s, with
$d\Phi_{\mathrm{ph}}/dt = -1.81\times10^{16}$ Mx/s ($\delta$ in
Fig.~\ref{fig:jet1tflux}), $d\overline{S}/dt = -8.25\times10^{13}$
cm$^{2}$/s ($\zeta$ in Fig.~\ref{fig:jet1tflux}), and $f=0.16$
($\theta$ in Fig.~\ref{fig:jet1bandc}) in the same manner.

From the integral of the emission increase around the K$_{2}$/K$_{1}$
on the Ca\emissiontype{II} K line profile (Fig.~\ref{fig:profiles1}c)
as a function of wavelength, the total flux from the the
Ca\emissiontype{II} K emission for the ``Jet09-0031'' was
$4.50\times10^{7}$ erg/s/cm$^{2}$.  The size of the bright core of the
``Jet09-0031'' was $4.90\times10^{15}$ cm$^{2}$ on the SOT
Ca\emissiontype{II} H broadband filter images.
So the estimated energy release rate of ``Jet09-0031'' by the
Ca\emissiontype{II} K emission becomes $2.20\times10^{23}$ erg/s.  The
radiative losses in the Ca\emissiontype{II} K line is about one order
of magnitude lower than the total energy loss in the visible spectrum
(\cite{avrett1986}, \cite{vernazza1981}).  Therefore the estimated
total energy release rate of ``Jet09-0031'' from the emission increase
will be of the order of $10^{24}$ erg/s.  This is comparable to the
magnetic energy release rate of $(1.1-8.1)\times10^{24}$ erg/s
estimated by a magnetic cancellation rate.

The durations of the magnetic cancellation at the footpoint of the two
jets (``Jet09-0045'' and ``Jet09-0031'') are both 130 seconds.  The
estimated released magnetic energies for the jets are
$(1.8-13)\times10^{26}$ erg and $(1.4-11)\times10^{26}$ erg,
respectively.
Released energy estimated by emission method gives value of the order
of $10^{26}$ erg for ``Jet09-0031'' during the life time of this jet,
270 seconds.

\subsection{Phenomenological model of typical chromospheric anemone jets}

In this section, we discuss the origin of typical jets described in
section~3 on the basis of observational data.

\subsubsection{Jet09-0031 and Jet09-0045}

\begin{figure*}
  \begin{center}
    \FigureFile(180mm,96mm){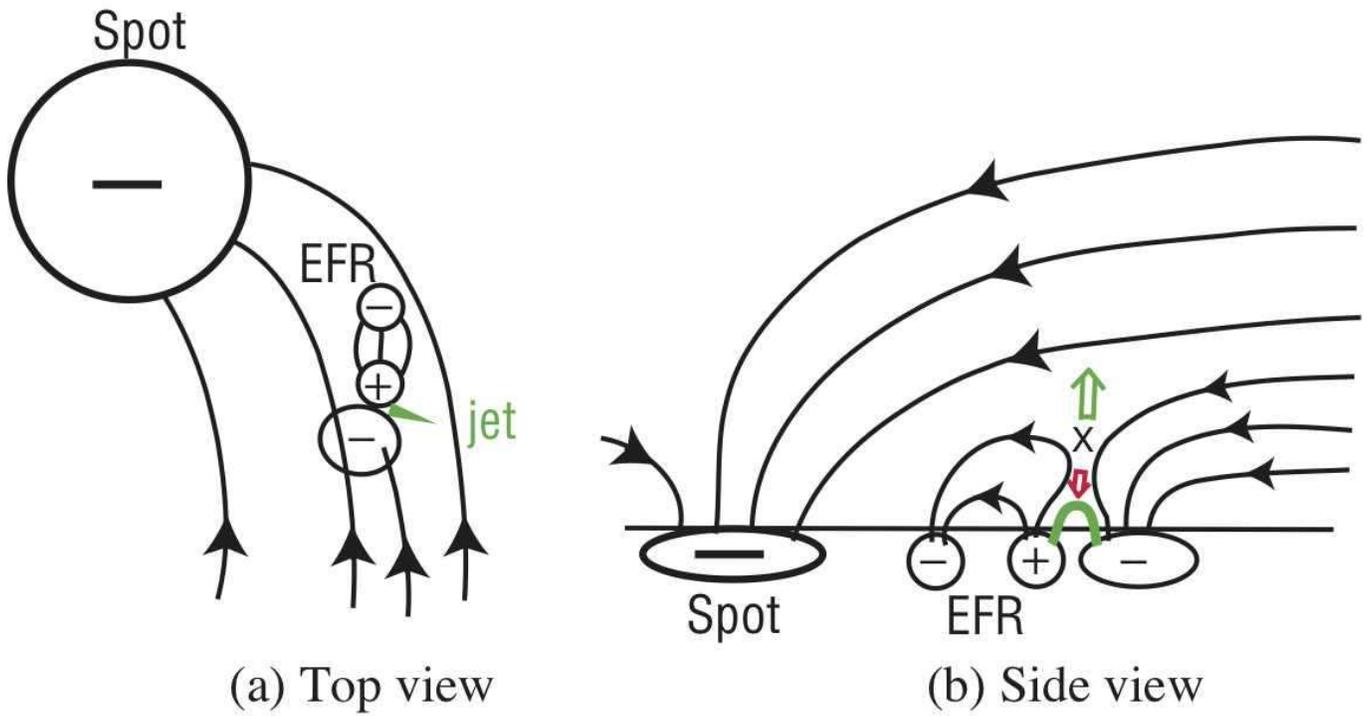}
  \end{center}
  \caption{A schematic diagram of ``Jet09-0031'' and ``Jet09-0045''.}\label{fig:cartoon1}
\end{figure*}

These jets occurred around 00:31--00:45 UT on Aug. 9, 2007.  The
magnetic field configuration during this time is illustrated in
Figure~\ref{fig:cartoon1}a.  This illustration is based on the
magnetogram, H$\alpha$ image, Ca\emissiontype{II}~H image, and
Ca\emissiontype{II} K spectroheliogram.  The jets occur at polarity
inversion line between emerging flux and pre-existing network patch.
This let us to think that magnetic reconnection occurred there.  Since
the jets and footpoint brightening were not observed in
Ca\emissiontype{II} K$_{3}$ images, the jets and reconnection must
have occurred below high or middle chromosphere.  How can we interpret
the blueshift (2--3 km/s) of K$_{3}$?  Since the jets were not seen in
the K$_{3}$ images and K$_{3}$ blueshift velocity is smaller than the
apparent jet velocities (5--10 km/s), the blueshift of K$_{3}$ may
represent the upward motion of dark filament seen in K$_{3}$ images
(Fig.~\ref{fig:dstbfinfi}c; see the H$\alpha$ images in
Fig.~\ref{fig:dstha1} for the dark filament), which may be triggered
by the reconnection.  On the other hand, the magnetic reconnection was
triggered by emerging flux, so that the reconnection point may not be
in the photosphere, but around lower or middle chromosphere.  So there
occurs a warm downflow that is heated by reconnection in the lower
chromosphere (Fig.~\ref{fig:cartoon1}b).  This downflow may be the
cause of the redshift of the emission increases in Ca\emissiontype{II}
K$_{1}$/K$_{2}$ component at the footpoint of the jet.
We note that, especially in the case of solar flares, one should be
careful about the interpretation of the blue or red asymmetries of
H$\alpha$ and Ca\emissiontype{II} K lines (e.g., \cite{ding1996},
\cite{ding1997}).  However this time, the observed red asymmetries can
be attributed to redshifts of K$_{2}$ emission since the redshifts are
obtained by measuring the bisector positions of the two flanks of the
K$_{2}$ emission, and the spectral range of the emission increases are
wide enough comparing to the blueshifts of the K$_{3}$ absorption
line.

\subsubsection{Jet08-0333}

\begin{figure*}
  \begin{center}
    \FigureFile(180mm,94mm){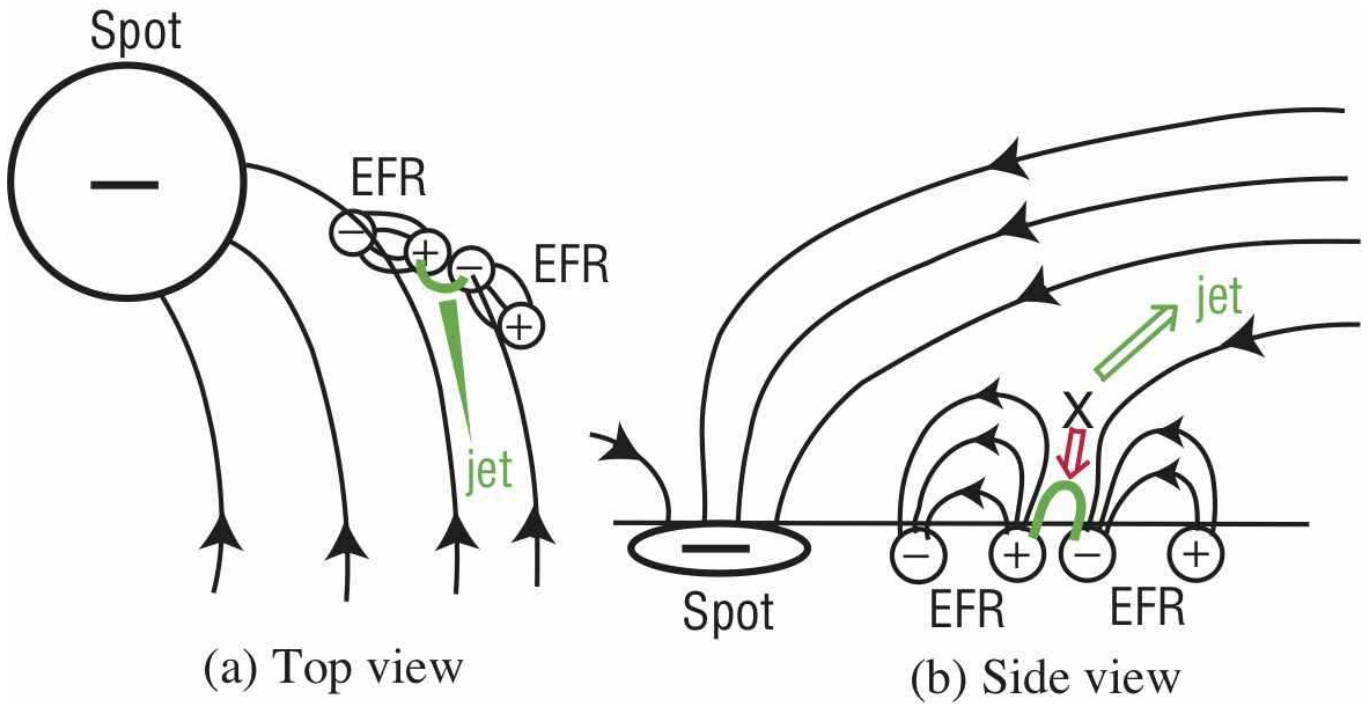}
  \end{center}
  \caption{A schematic diagram of ``Jet08-0333''.}\label{fig:cartoon2}
\end{figure*}

This jet occurred around 03:33 UT on Aug. 8, 2007, one day before the
above jets. In this case, the jet was not seen in K$_{3}$ image, while
blue shift was observed in K$_{3}$. The jet or brightening was
observed in K$_{1}$ and K$_{2}$ images near the polarity inversion
line between two emerging flux (Fig.~\ref{fig:cartoon2}a,
\ref{fig:cartoon2}b). Hence the reconnection may have occurred in the
low or middle chromosphere, so that the downflow may cause the
redshift of the emission increases in Ca\emissiontype{II}
K$_{1}$/K$_{2}$ component.  The overall magnetic field configuration
and the associated dynamics may be similar to those for Jet09-0031.

\subsubsection{Jet08-0046}

\begin{figure*}
  \begin{center}
    \FigureFile(180mm,96mm){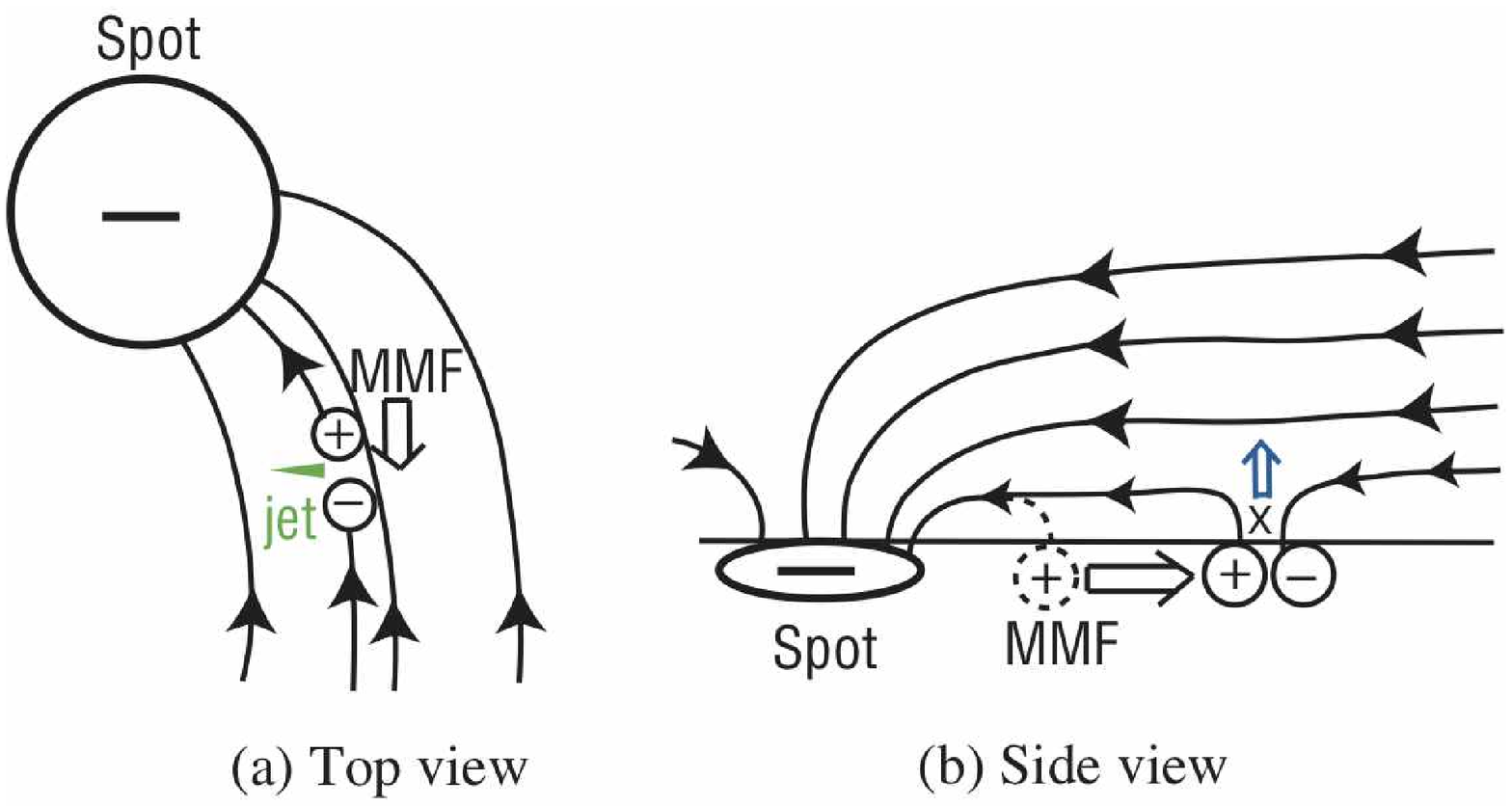}
  \end{center}
  \caption{A schematic diagram of ``Jet08-0046''.}\label{fig:cartoon3}
\end{figure*}

The environment of this jet is different from those of previous
cases. In this case, the jet was triggered by moving magnetic feature
(Fig.~\ref{fig:cartoon3}a, \ref{fig:cartoon3}b). Hence the
reconnection may occur in the photosphere, so that we can observe only
the upward flow in lower chromosphere, which may cause the blueshift
of the emission increases in Ca\emissiontype{II} K$_{1}$/K$_{2}$
component.

\subsection{Relation to Other Jets and Jet-like Phenomena}

In this section we compare our results with previous observations.
\citet{kubota1974} reported red asymmetry in Na D line at the root of
the surges.  This may be similar to red asymmetry of
Ca\emissiontype{II} K$_{1}$/K$_{2}$ at the footpoint of the
chromopsheric anemone jets.  Such downflow may correspond to
reconnection jet ejected downwards from the reconnection point.

Our observations of chromospheric anemone jets has revealed that the
jets are not seen in Ca\emissiontype{II} K$_{3}$ but are bright in
K$_{1}$ and K$_{2}$.  This is similar to the characteristics of
Ellerman bombs in H$\alpha$ line (\cite{ellerman1917},
\cite{roy1973b}, \cite{kitai1983}, \cite{qiu2000},
\cite{geolgoulis2002}, \cite{pariat2004}, \cite{fang2006},
\cite{matsumoto2008a}, \yearcite{matsumoto2008b},
\cite{watanabe2008}).  Recently, \citet{pariat2007} observed Ellerman
bombs with Ca\emissiontype{II} 8542 \AA\ line, and found that
Ca\emissiontype{II} 8542 \AA\ line show similar ``moustache'' like
intensity profile as in the H$\alpha$ line.  So it is likely that the
footpoint of the chromospheric anemone jets reported previously
correspond to the Ellerman bombs.  It is interesting to note that many
Ellerman bombs show elongated structure \citep{kurokawa1982} and often
become the root of surges (\cite{roy1973a}, \cite{roy1973b}).
\citet{geolgoulis2002} and \citet{pariat2004} proposed the magnetic
reconnection model for Ellerman bombs in which the reconnection occurs
in the sea-serpent flux tubes during the resistive emergence of
magnetic flux.  \citet{isobe2007} corroborated this model using 2D
magnetohydrodynamic numerical simulations of emerging flux.
\authorcite{matsumoto2008a}(\yearcite{matsumoto2008a},
\yearcite{matsumoto2008b}) and \citet{watanabe2008} obtained the
observational support of this model using H$\alpha$ line profile
analysis based on the spectroscopic observations at Hida Observatory.

If our interpretation of Ellerman bombs as the footpoint of the
chromospheric anemone jets is correct, then our observations on the
jets associated with emerging flux (EFR) and moving magnetic features
(MMFs) give a model that is different from the previous model
(\cite{geolgoulis2002}, \cite{pariat2004}, and \cite{isobe2007}).  In
our model, the magnetic reconnection is triggered by the collision of
the EFR or the MMF with the pre-existing magnetic flux.  The
association of Ellerman bombs with MMFs has been observed by
\citet{nindos1998}.  The common point in these models is that the
height of magnetic reconnection is low in the atmosphere, either the
lower chromosphere or the photosphere.  So the numerical simulation of
reconnection in the lower chromosphere and/or the photosphere (e.g.,
\cite{takeuchi2001}, \cite{chen2001}, \cite{isobe2008}) should be
developed further.

A typical energy released in the Ellerman bombs has been estimated to
be of the order 10$^{26}$--10$^{27}$ erg \citep{fang2006}.  Based on
our observation, the total energy of the chromospheric anemone jets is
estimated to be of the order 10$^{26}$ erg, which is smaller than that
of Ellerman bombs. This suggests that even if the basic physics of
chromospheric anemone jets is similar to that of classical Ellerman
bombs, the total energy and size of the chromospheric anemone jets
(especially smaller ones) are smaller than those of Ellerman bombs.
This collaborate the idea of \citet{shibata2007} that there may be
even smaller non-resolved reconnection events (or nanoflares) which
may play an important role in heating the chromosphere and corona
\citep{parker1988}.

\bigskip
Acknowledgement

This work is supported in part by the Grant-in-Aid for Creative
Scientific Research ``The Basic Study of Space Weather Prediction''
(Head Investigator: K. Shibata) from the Ministry of Education,
Culture, Sports, Science, and Technology of Japan, and in part by the
Grand-in-Aid for the Global COE program ``The Next Generation of
Physics, Spun from Universality and Emergence'' from the Ministry of
Education, Culture, Sports, Science, and Technology (MEXT) of Japan.
Hinode is a Japanese mission developed and launched by ISAS/JAXA,
collaborating with NAOJ as a domestic partner, NASA and STFC (UK) as
international partners. Scientific operation of the Hinode mission is
conducted by the Hinode science team organized at ISAS/JAXA. This team
mainly consists of scientists from institutes in the partner
countries. Support for the post-launch operation is provided by JAXA
and NAOJ (Japan), STFC (U.K.), NASA, ESA, and NSC (Norway).
SM is grateful to H. Isobe for his valuable comments.  SM is grateful
to K.A.P. Singh and S. Yashiro for their English corrections.

\appendix

\section{H$\alpha$ Observations}

\subsection{Jet09-0031 \& 0045}

\begin{figure*}
  \begin{center}
    \FigureFile(180mm,47mm){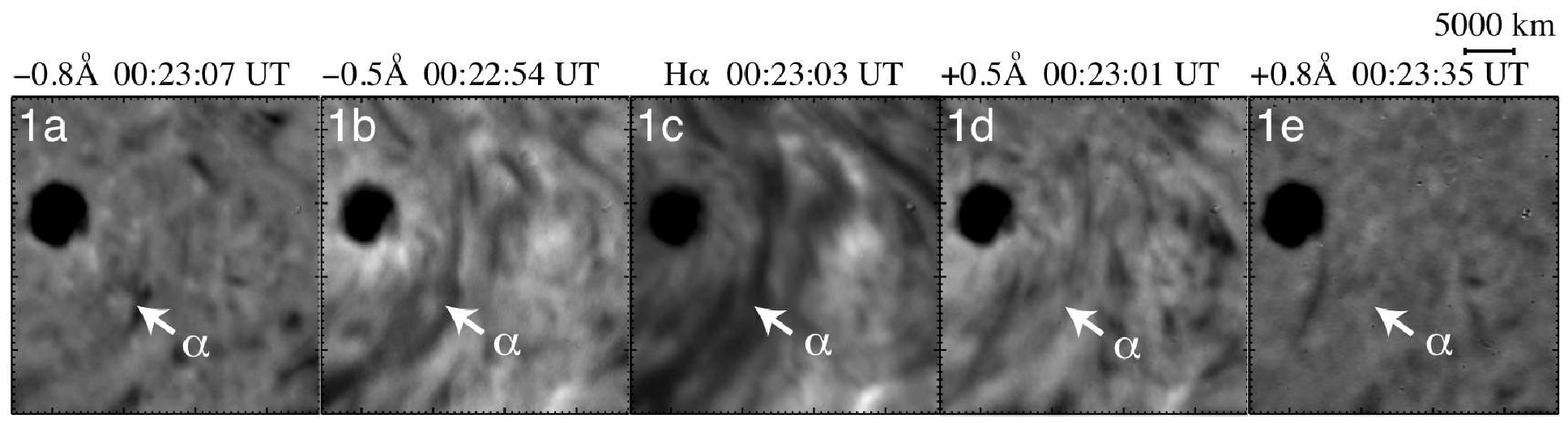}
    \FigureFile(180mm,47mm){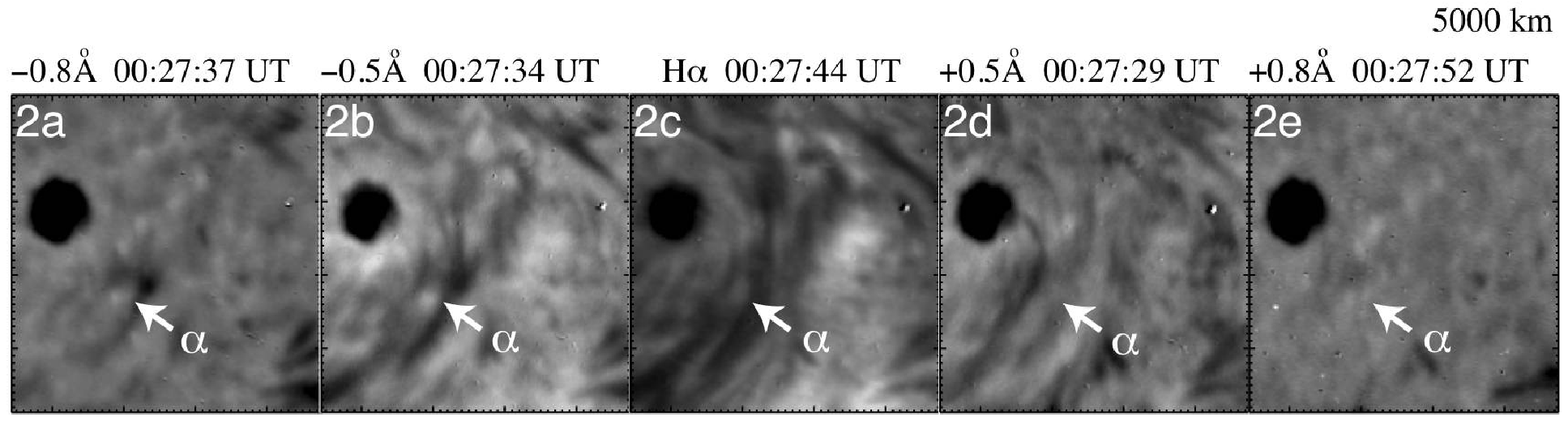}
    \FigureFile(180mm,47mm){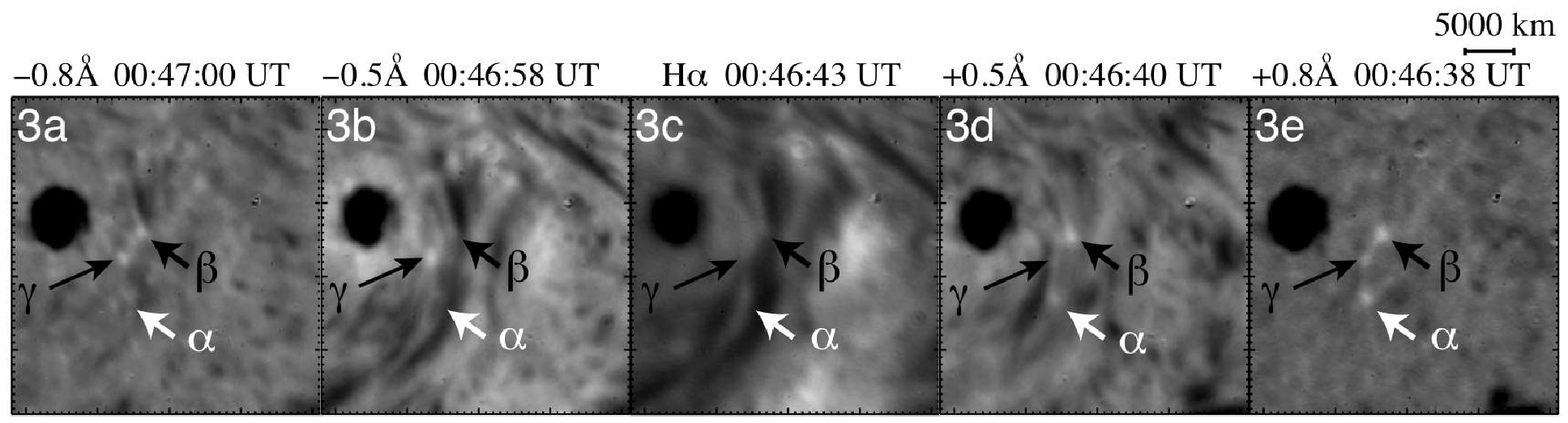}
  \end{center}
  \caption{Three sets of snapshots of the series of jet events on
    August~9 around 00:45~UT with DST/Hida H$\alpha$ Lyot filter
    images at five wavelengths positions (H$\alpha$ center and its
    wing at $\pm$0.5\AA\ and $\pm$0.8\AA): Top is north and left is
    east.  Each image has the same FOV of
    43.\arcsec2$\times$43.\arcsec2, the same FOV of
    Fig.~\ref{fig:dstbfinfi}.  The white arrows ``$\alpha$'' indicate
    the locations of the jet bright cores in Ca\emissiontype{II}~H
    images. The upper row (1a--1e) represents the time of 4 minutes
    before the 2nd row of Fig.~\ref{fig:evoljet1} (``Jet09-0031'').
    No obvious feature is seen in H$\alpha$ at that time.  The middle
    row (2a--2e) represents the time of the 2nd row of
    Fig.~\ref{fig:evoljet1} (``Jet09-0031'').  The bottom row (3a--3e)
    represents the time of the Ca\emissiontype{II}~H intensity peak of
    the strongest jet (``Jet09-0045'') in this series of jet events
    (cf. 3rd row of Fig.~\ref{fig:evoljet1}).  The bright H$\alpha$
    counterpart of this jet is seen in the red wing of H$\alpha$
    (shown arrows ``$\alpha$'' in 2d and 2e).  The locations of the
    other bright points ``$\beta$'' and ``$\gamma$'' correspond to
    another small chromospheric jets that occurred around 00:40~UT and
    00:30~UT, respectively.}\label{fig:dstha1}
\end{figure*}

Figure~\ref{fig:dstha1} shows three sets of snapshots of H$\alpha$
Lyot filter images at five wavelength positions (H$\alpha$ center and
its wing at $\pm$0.5\AA\ and $\pm$0.8\AA) for this series of jets with
DST/Hida.  Each image has the same field of view of
Fig.~\ref{fig:dstbfinfi}.  We use the H$\alpha$ images for examining
the relation between the chromospheric anemone jets and Ellerman
bombs.  The corresponding locations for the jet bright cores from the
SOT Ca\emissiontype{II} H broadband filter images are pointed by
arrows ``$\alpha$'' in the figure.

No obvious feature is seen in the H$\alpha$ Lyot filter images at the
five wavelength positions at the time of 8 minutes before the
``Jet09-0031'' Ca\emissiontype{II} H intensity peak (see 1a--1e; 4
minutes before the time of the jet onset frame (2a) in
Fig.~\ref{fig:evoljet1}).  Around the time of the intensity peak of
``Jet09-0045'' (3a--3e), the bright H$\alpha$ counterpart of this jet
is seen in the red wing of H$\alpha$ (see the arrows ``$\alpha$'' in
2d and 2e).  The locations of the other bright points ``$\beta$'' and
``$\gamma$'' correspond to another small jets that occurred around
00:40 UT and 00:30 UT, respectively.  The time of the middle row
(2a--2e) is chosen similar to for Fig.~\ref{fig:dstbfinfi}, to
understand the global magnetic field orientations in the chromosphere
around the jets.  The orientations of the dark filaments and fibrils
around the jet region were approximately from the north-northwest to
the south-southeast.  It is approximately in the same direction as
that of the series of the centers of the magnetic elements that
involved EFR.

\subsection{Jet08-0333}

\begin{figure*}
  \begin{center}
    \FigureFile(180mm,59mm){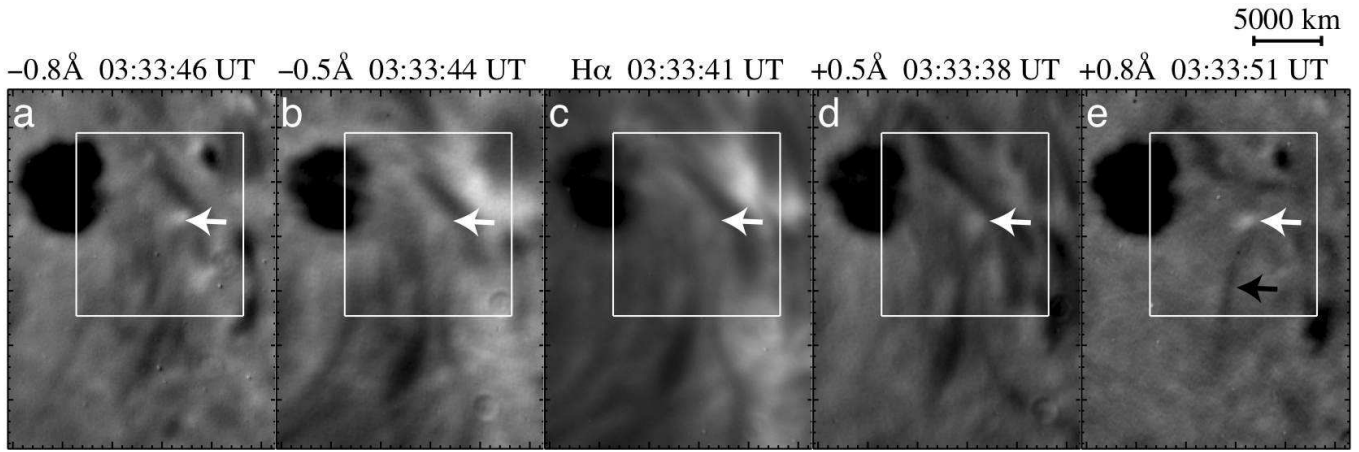}
  \end{center}
  \caption{A set of snapshots of ``Jet08-0333'' with DST/Hida
    H$\alpha$ Lyot filter images at five wavelengths positions (H$\alpha$
    center and its wing at $\pm$0.5\AA\ and $\pm$0.8\AA): Top is north
    and left is east.  Each image has the same FOV of
    16.\arcsec2$\times$19.\arcsec8.  The white boxes represent the FOV
    of Fig~\ref{fig:evoljet2}.  The time of these images are chosen as
    around the time of the Ca\emissiontype{II} H intensity peak of
    this jet, and corresponds to the 4th row of
    Fig.~\ref{fig:evoljet2}.  The white arrows indicate the location
    of the jet bright core in SOT Ca\emissiontype{II} H images.  The
    bright H$\alpha$ counterpart of the jet bright core is seen in the
    wing of H$\alpha$, especially in (e), at $-$0.8\AA\ of the
    H$\alpha$ line center.  The black arrow in (e) indicates a surge,
    which appeared around this time and has the coincide location of
    the chromospheric anemone jet.}\label{fig:dstha2}
\end{figure*}

Figure~\ref{fig:dstha2} shows a set of snapshots of ``Jet08-0333'' at
its SOT Ca\emissiontype{II} H intensity peak time with DST/Hida
H$\alpha$ Lyot filter images at five wavelength positions (H$\alpha$
center and its wing at $\pm$0.5\AA\ and $\pm$0.8\AA).  The white boxes
represent the FOV in Fig~\ref{fig:evoljet2}.  The white arrows
indicate the location of the jet bright core in Ca\emissiontype{II} H
broadband filter images.  The bright H$\alpha$ counterpart of the jet
bright core is seen in the wing of H$\alpha$, especially at
$-$0.8\AA\ of the H$\alpha$ line center.  The black arrow in
Fig.~\ref{fig:dstha2}e indicates a surge, which appeared around the
peak time of this jet, in the same location of the jet in
Ca\emissiontype{II} H broadband filter images, but the length of the
surge was longer.  Mass motions of this surge was southward, same as
``Jet08-0333''.  According to the shapes of fibrils and dark filaments
surrounding jet region, the directions of the global magnetic field
lines in the chromosphere around the ``Jet08-0333'' were southward.

\subsection{Jet08-0046}

\begin{figure*}
  \begin{center}
    \FigureFile(180mm,59mm){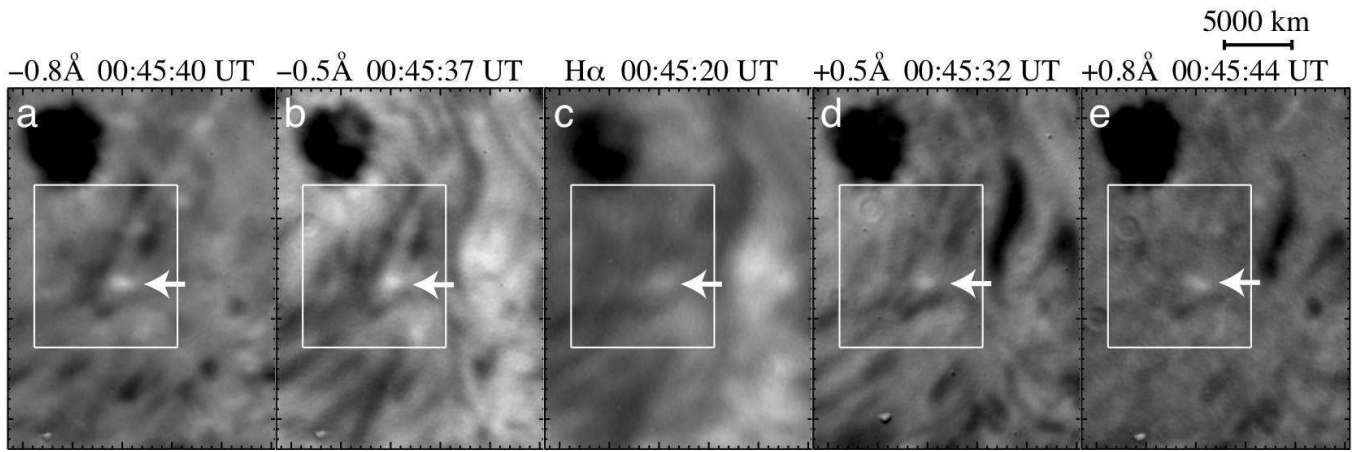}
  \end{center}
  \caption{A set of snapshots of ``Jet08-0046'' with DST/Hida
    H$\alpha$ Lyot filter images at five wavelengths positions (H$\alpha$
    center and its wing at $\pm$0.5\AA\ and $\pm$0.8\AA): Top is north
    and left is east.  Each image has the same FOV of
    27\arcsec$\times$36\arcsec.  The white boxes represent the FOV of
    Fig~\ref{fig:evoljet3}.  The time of these images are chosen as
    around the time of the Ca\emissiontype{II} H intensity peak of
    this jet, and corresponds to the 4th row of
    Fig.~\ref{fig:evoljet3}.  The white arrows indicate the location
    of the jet bright core in SOT Ca\emissiontype{II} H images.  In
    this ``Jet08-0046'', the bright H$\alpha$ counterpart of the jet
    bright core is widely seen in the all five wavelengths positions but
    faint at the H$\alpha$ line center.}\label{fig:dstha3}
\end{figure*}

Figure~\ref{fig:dstha3} shows set of snapshots of ``Jet08-0046'' at
its SOT Ca\emissiontype{II} H intensity peak time with DST/Hida
H$\alpha$ Lyot filter images at five wavelength positions (H$\alpha$
center and its wing at $\pm$0.5\AA\ and $\pm$0.8\AA).  The white boxes
represent the FOV in Fig~\ref{fig:evoljet3}.  The white arrows
indicate the location of the jet bright core in SOT
Ca\emissiontype{II} H broadband filter images.  The white arrows
indicate the location of the jet bright core in Ca\emissiontype{II} H
images.  In this ``Jet08-0046'', the bright H$\alpha$ counterpart of
the jet bright core is widely seen in all the five wavelength
positions but faint at the H$\alpha$ line center.  According to the
shapes of fibrils and dark filaments in this area, the directions of
the global magnetic field lines in the chromosphere around the
``Jet08-0046'' were in the southeast direction.

\section{Detailed evolution of a single jet ``Jet09-0045''}

\begin{figure*}
  \begin{center}
    \FigureFile(180mm,91mm){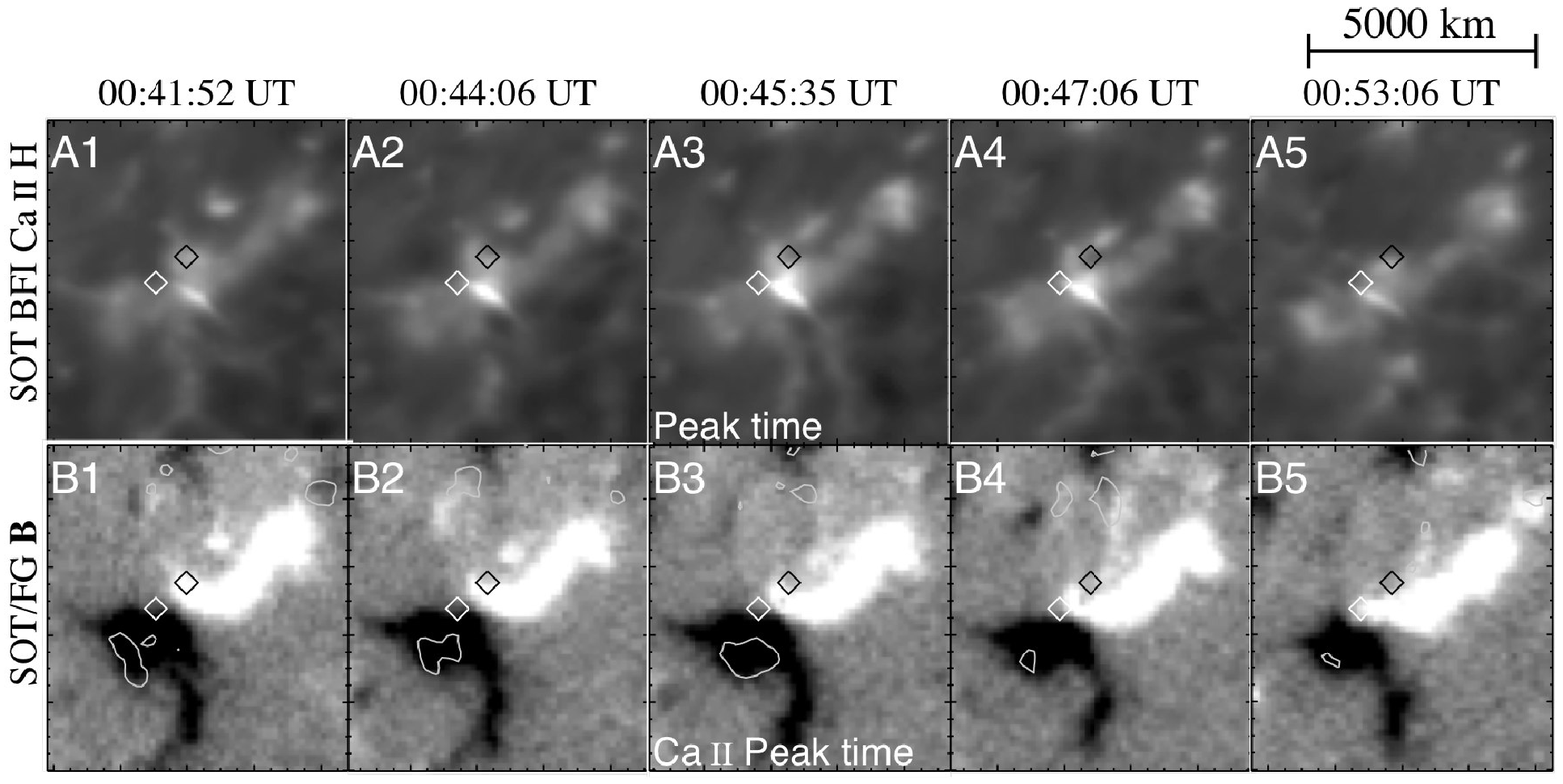}
  \end{center}
  \caption{Detailed evolution of a single jet ``Jet09-0045'', in the
    series of jet events shown in Fig.~\ref{fig:evoljet1}, with the
    SOT BFI Ca\emissiontype{II}~H (upper panels) and with the SOT/FG
    magnetograms (lower panels): Top is north and left is east.  Each
    image has the same FOV of 9\arcsec$\times$9.\arcsec6.  In the
    magnetograms, the white and black gray scale color shows the
    positive and negative line-of-sight component of the photospheric
    magnetic field {\bf B} (color is saturated at $\pm$150 Gauss),
    while the contours show the horizontal components of {\bf B}
    (contour level: 200 Gauss).  The diamonds show the footpoints
    locations of the bright cusp of this jet at the time of the
    Ca\emissiontype{II}~H intensity peak (see A3).  This time series
    corresponds to the strongest single peak of the
    Ca\emissiontype{II}~H light curve, and the same timing of the
    largest magnetic flux cancellation (cf. Fig.~\ref{fig:jet1tflux}).
  }\label{fig:evoljet1detailed}
\end{figure*}

Figure~\ref{fig:evoljet1detailed} shows a detailed evolution of a
single jet ``Jet09-0045'' of the series of jets, with SOT
Ca\emissiontype{II}~H broadband filter images and with SOT/FG
magnetograms.  This time series corresponds to the strongest single
peak of Ca\emissiontype{II}~H intensity curve, and the same timing of
that the largest magnetic flux cancellation around the polarity
inversion line occurred (cf. $\gamma$ in Fig.~\ref{fig:jet1tflux}).
The diamonds show the footpoints locations of the bright cusp of this
jet at the time of the Ca\emissiontype{II}~H intensity peak (frame
A3).  It is obvious that the footpoints of the jet (``Jet09-0045'')
was the most actively colliding area between the two opposite polarity
magnetic sources.  The positive polarity magnetic source, which is
involved in an EFR in the north, was actively entering into the
isolated negative source from its northwest.  Then a part of this
negative source was, seemingly, ``disappearing''.  The bright cusp of
the jet appeared in this timing.


\end{document}